\documentclass[twocolumn]{aastex631}

\shorttitle{The AH Pictoris syndrome}
\shortauthors{Albert Bruch}

\begin{document}

\title{The AH Pictoris syndrome: Continuous trains of stunted outbursts
in novalike variables}

\author{Albert Bruch}
\affiliation{Laborat\'orio Nacional de Astrof\'{\i}sica \\
Rua Estados Unidos, 154, \\
37504-364 Itajub\'a -- MG, Brazil}

\begin{abstract}
Novalike variables are a subgroup of cataclysmic 
variables (CVs) that -- unlike dwarf novae -- 
do not exhibit strong brightenings in their long-term light curves. 
Variations over time scales of weeks, months or years are mostly restricted 
to irregular low-amplitude modulations. However, some of 
them occasionally suffer from so-called stunted outbursts, that is, small-scale
brightenings of less than a magnitude lasting for a couple of days to weeks.
There is no consensus about the physical mechanisms behind these
outbursts. Here I discuss the common properties
of a group of novalike variables (which I call AH~Pictoris stars
after its most prominent member) that exhibit a continuous train of 
successive stunted outbursts over their entire observational history, or at 
least for several years. The outburst amplitudes are stable in a given 
system, always ranging between 0.5 and 1~mag in the visual band. The 
outburst intervals, at an overall range between 12 and 30 days, and the
outburst profiles can gradually evolve, but no sudden 
changes are observed. On shorter time scales the 
orbital waveforms are not only surprisingly similar, 
but also evolve in the same way over the outburst cycle. All AH~Pic
stars have absolute visual magnitudes in the overlap region between absolute
magnitudes of all novalike variables and of quiescent dwarf novae above the
CV period gap. So far, I
identified seven novalike variables with the consistent photometric 
behavior which may be termed the AH~Pic syndrome. Several more systems may
be related objects. The relationship to the anomalous Z~Cam stars is 
discussed.
\end{abstract}


\section{Introduction}
\label{Introduction}

As is well known since the development, in the 1960ies and 70ies, of a basic 
model, summarized in the excellent early reviews by \citet{Warner76} and
\citet{Robinson76},
cataclysmic variables (CVs) are interacting binary systems where a
late type secondary star -- in most cases on or close to the main sequence --
transfers matter to a white dwarf primary. Before the matter can settle
on the surface of the compact object, it forms an accretion disk around
the white dwarf if no strong magnetic fields are present, or otherwise 
forms accretion columns or curtains, sometimes in combination with truncated 
disks.

Non-magnetic CVs can roughly be divided into dwarf novae that undergo 
semi-regular outbursts with amplitudes of $\approx$2--6~mag and intervals
ranging from days to years, and novalike variables (NLs) that do not suffer 
from such eruptions but remain on a rather stable long-term brightness level
(except for occasional low states in the subclass of VY~Scl stars). Here,
I also count among the NLs most classical or recurrent novae because long 
after the large-scale nova outbursts which occur in intervals of decades
to millennia in these systems they behave much like NLs.

The reason for the distinct long-term photometric behavior or dwarf novae
and NLs is a much lower average mass transfer rate from the secondary star
in the former. The accretion disk is then in a low viscosity state and
transfers less matter to the white dwarf than it receives from the late type
component. Therefore, material is accumulated until a disk instability 
mechanism (DIM) sets in \citep[for a review, see][]{Lasota01} that elevates
the disk viscosity so that the excess matter is rapidly dumped upon the
primary star. This is seen as a dwarf nova outburst. Once the disk is
depleted of mass, it returns to its low viscosity state, the outburst ends,
and the cycle begins again. The DIM only works if the average mass transfer
rate remains beneath a critical level $\dot{M}_{\rm crit}$. In novalike 
variables it is above this limit which means that the accretion disk is
always in a high viscosity state. The mass accretion rate onto the white
dwarf is equal to the mass transfer rate, the disk is in a steady state, and
dwarf nova type eruptions are suppressed. In a sense NLs can thus be regarded
as dwarf novae in a permanent outburst.

Of course, this is an idealized picture and should be taken {\it cum granum 
salis}! Even disregarding the low states of VY~Scl stars, NLs suffer from
moderate long-term brightness variations \citep[see, e.g., the recent synoptic 
study of eclipsing novalike variables of][]{Bruch24a}. Since the brightness 
of CV is to a large part determined by the mass transfer rate within the
accretion disk, and therefore -- in NLs -- by the rate at which the disk 
receives matter from the secondary star, modulations of their brightness 
can at least in part be explained by
variations in the mass transfer rate. But other brightness variations
may require different mechanisms, including some kind of modified DIM. Here,
I am referring to small scale (amplitude: some tenths of a magnitude) 
brightenings, lasting for about one or two weeks observed in some NLs. They 
were first described in some detail by \citet{Honeycutt98a} who coined the term 
`stunted outbursts' for these events. 

Such stunted outbursts occur in general irregularly and sparsely. However,
a small number of stars are known to exhibit continuous trains of such
event over years. These are CM~Del \citep{Zsidi23}, 
FY~Per \citep{Honeycutt04}, V1116~Cep
\citep{Shears20, Shears23}, and KIC~9202990 \citep{Ostensen10, Ramsay16}.
Some more systems are identified in this contribution. I call them AH~Pic 
stars after the prototype that exhibits their peculiar photometric behavior 
most clearly. In the above mentioned papers, the respective stars were always 
described as isolated cases with unusual properties. They were never 
discussed as a group. The purpose of the present paper is to draws attention 
to the commonalities between the members of this group and to review existing
ideas (and their problems) concerning the origin of stunted outbursts.

In Sect.~\ref{The data} I
present the sources for the data which form the basis of this study. 
Sect.~\ref{Target stars and light curves} starts with a brief description of 
the common properties of the AH~Pic stars. Subsequently, a characterization 
of the individual stars identified as belonging to this group, and of their 
photometric behavior is given. An analysis of the long-term properties,
the absolute magnitudes, and the orbital variations of the AH~Pic stars 
then follows in Sect.~\ref{Properties of the AH Pic stars}.
Stars with stunted outbursts that are possibly related to AH~Pic stars are
reviewed in Sect.~\ref{Related systems}. A general discussion and a short
summary conclude this work in Sects.~\ref{Discussion} and \ref{Summary}.

\section{The data}
\label{The data}

This study is exclusively based on archival data. Light curves of all but one
of the target
stars were observed by the Transiting Exoplanet Survey Telescope (TESS) 
\citep{Ricker14} and were downloaded from the Barbara A.\ Mikulski Archive for
Space Telescopes (MAST)\footnote{https://archive.stsci.edu}. They have a
time resolution of 120~sec. For a more
detailed description of their structure and handling in the present context,
see \citet{Bruch23b}. In brief, I only use Simple Aperture Photometry 
(SAP) data because the Pre-Search Data Conditioning Simple Aperture Photometry 
(PDCSAP) process tends to severely distort variations on time scales
longer than a few days which are crucial for this study. Data of the same
objects taken in overlapping TESS sectors observed in immediate succession 
are combined into single light curves (referred to as LC\#1, LC\#2
etc.).

One of the stars, KIC~9202990, was also observed over an extended period of
time as part of the Kepler space mission \citep{Borucki10} in long cadence
mode with a time resolution of 29.4~min. During some limited intervals
also light curves taken in short cadence mode with a time resolution of 
59~sec are available. Both data sets (again, SAP data) are used here. They
were also downloaded from MAST.

Additionally, all targets have been observed regularly during the past 10 
years or so by the All-Sky Automated Survey for Supernovae (ASAS-SN) 
\citep{Kochanek17}. They were downloaded from the ASAS-SN Sky Patrol 
website\footnote{https://asas-sn.osu.edu}. While the survey aims to 
observe the entire sky once every day, even apart from
seasonal gaps additional interruptions occur. Thus, the observing cadence is
somewhat irregular. In the early years observations were taken in the $V$ band
while in more recent years the $g$ band is employed.

Finally, light curves extending over several years were downloaded from the 
American Association of Variable Star Observers (AAVSO) International 
Database\footnote{https://www.aavso.org/data-download} \citep{Kafka21}.
They complement the ASAS-SN data of three targets (V2837~Ori, FY~Per and
V1116~Cep). Only data observed in the $V$ band (or reduced to the 
$V$ band) are used.

\section{The AH-Pic syndrome, identified group members, and light curves}
\label{Target stars and light curves}

The defining property of AH~Pic stars is a continuous sequence of small 
scale brightenings recurring on time scales of a couple of weeks with no or 
only small quiescent inter-outburst phases between them. As will be detailed
below, this behavior is either 
permanent, or active phases over longer time intervals (years) are interrupted 
by periods without this kind of activity. The intervals between brightenings
are not constant, but evolve slowly over various cycles, such that longer and
shorter intervals are grouped but not intermingled. The outburst shape also
evolves slowly between outbursts.
Other common properties of the AH~Pic stars are outburst amplitudes between
0.5 and 1~mag, surprisingly similar orbital waveforms that change
consistently between brighter and fainter phases, and absolute visual 
magnitudes in the overlap between the absolute magnitude
distributions of all NLs and of dwarf novae above the CV period gap.
Collectively, I refer to
the peculiar properties of these stars as the AH~Pic syndrome.

All objects identified as AH~Pic stars are classified in widely
used catalogues such as \citet{Downes01} or \citet{Ritter03} and/or in the
few more detailed publications on the specific systems (see below) as
novalike variables. However, this classification is not unambiguous as
other sources assign them to other possible classes. In particular, in the
International Variable Star Index\footnote{https://www.aavso.org/vsx/}
they are also classified as UG (dwarf nova), UGZ (Z~Cam-type dwarf nova) 
or IW (anomalous dwarf nova; IW~And-type star). The relationship of the
AH~Pic stars in particular to the IW~And-type stars will be explored in
more detail in Sect.~\ref{Relationship to anomalous Z Cam stars}.
  
\begin{figure*}
\figurenum{1}
\label{ahpic-lc}
\plotone{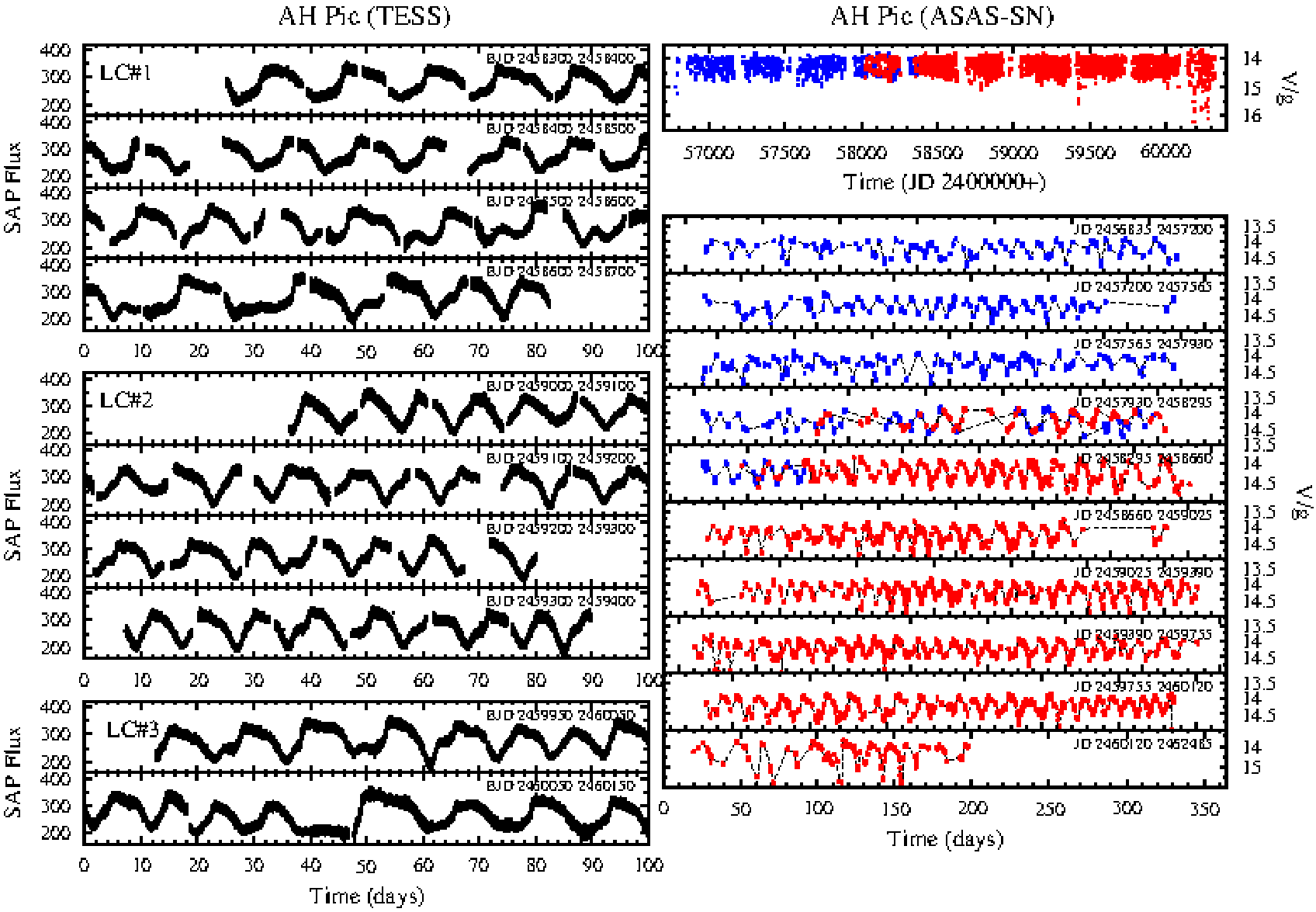}
\caption{Light curves of AH Pic. On the left TESS data taken in 
sectors 1--13, 27--39 (with a gap corresponding to sector 36), and sectors
61--67, respectively, are shown. The upper panel on the right contains the
full $\sim$10~yr ASAS-SN light curve, while the other panels show the 
seasonal light curves on an expanded time scale. Blue and red dots refer
to the $V$ and $g$ bands, respectively. Thin broken lines between the data
points are meant to guide the eye.}
\end{figure*}

The most striking and obvious manifestation of the AH~Pic syndrome
is observed in
two long TESS light curves of AH~Pic which both span about a year and are
separated by a 12 months interval, plus an additional shorter light curve. 
They are shown in Fig.~\ref{ahpic-lc} (left panel). The coverage by 
TESS of the other stars was less intensive, but for all objects light curves
from other sources (either/or Kepler, ASAS-SN or AAVSO)
confirm their common properties. 

So far, I could identify seven NLs which exhibit at least during
extended periods of time (years) the continuous sequence of stunted 
outbursts that is the principle manifestation of the AH~Pic syndrome.
Their basic parameters and the main results of this study
are summarized in Table~\ref{Table: AH Pic stars}.
Before characterizing the AH~Pic syndrome in more detail, I 
briefly present the seven target stars and the light curves available for
analysis. I start with the prototype AH~Pic.

%--------------------------------------------------------------
\begin{table*}
\caption{Summary of the properties of AH~Pic stars}
\label{Table: AH Pic stars}
\centering

\begin{tabular}{llllllll}
\hline \hline
Property & V1116 Cep & CM Del & PS Eri & V2837 Ori & FY Per & AH Pic & KIC 920229  \\
\hline
orb.\ Period [d]    & 
0.161439(9)$^c$     &
0.141052(3)$^a$     &
\ldots              &
0.27610702(2)$^b$   &
0.25837(7)$^c$      &
0.1409122(2)$^a$    &
0.1659404$^d$       \\
$V^e$               &
$14.25(16)$         &
$14.14(10)$         &
$13.65(16)$         &
$14.99(16)$         &
$12.80(13)$         &
$14.61(15)$         &
$15.42(19)$         \\
distance (pc)$^f$   &
$487(3)$            &
$393(3)$            & 
$481(3)$            &
$650(10)$           &
$342(2)$            &
$570(5)$            &
$985(18)$           \\
$E(B-V)^g$          & 
$0.076$             &
$0.030$             &
$0.010$             &
$0.068$             &
$0.128$             &
$0.024$             &
$0.034$             \\
$M_V$               & 
$5.57(23)$          &
$6.07(16)$          &
$5.21(16)$          &
$5.71(33)$          &
$4.73(23)$          &
$5.76(15)$          &
$5.35(21)$           \\
light curve source$^h$     &
T/AS/AA             &
T/AS                &
AS                  &
T/AS/AA             &
T/AS/AA             &
T/AS                &
K/T/AS              \\
median amplitude (mag)  &
%                    &
                    &
                    &
                    &
                    &
                    &
                    \\
\phantom{mm}TESS/Kepler  &
\ldots              &
\ldots              &
\ldots              &
\ldots              &
\ldots              &
0.45                &
0.58                \\
\phantom{mm}ASAS-SN/AAVSO  &
0.59                &
0.51                &
0.95                &
0.98                &
0.65                &
0.66                &
0.83                \\
rel.\ width (ampl.\ distr.)  &
                    &
                    &
                    &
                    &
                    &
                    &
                    \\
\phantom{mm}TESS/Kepler  &
\ldots              &
\ldots              &
\ldots              &
\ldots              &
\ldots              &
0.12                &
0.17                \\
\phantom{mm}ASAS-SN/AAVSO  &
0.24                &
0.24                &
0.17                &
0.19                &
0.19                &
0.24                &
0.24                \\
median interval (d) &
13.9                &
17.9                &
16.3                &
30.0                &
25.1                &
11.9                &
13.0                \\
rel.\ width (interval distr.)  &
0.35                &
0.17                &
0.13                &
0.17                &
0.24                &
0.22                &
0.23                \\
\hline
\end{tabular}
\\ 
$^a$: \citet{Bruch23b};
$^b$: \citet{Bruch24b};
$^c$: \citet{Bruch24a};
$^d$: \citet{Ramsay16};
$^e$: average ASAS-SN magnitude during minima (disconsidering dips);
$^f$: \citet{Bailer-Jones21};
$^g$: \citet{Capitanio17};
$^h$: T=TESS; K=Kepler; AS=ASAS-SN; AA=AAVSO
\end{table*}
%------------------------------------------------------------------

\subsection{AH Pictoris}
\label{AH Pic}

Only few details are known about \object{AH Pic} (= EC~05565-5934). The star 
was identified as a CV by \citep{Chen01} in the Edinburgh-Cape Survey and
classified as a UX~UMa-type cataclysmic variable. Their photometry suggested
an orbital period of 0.142(2)~d. This was confirmed spectroscopically
by \citet{Rodriguez-Gil07} who measured a period of 0.146(4)~d. They
classify AH~Pic as a SW~Sex star and estimate a low orbital inclination
of $25\degr < i < 37\degr$. More recently, based on TESS data \citet{Bruch24b} 
measured a more precise period of 0.1409122(2)~d. 

TESS observed AH~Pic extensively in sectors 1--13, 27--35, 37--39, and 
61--67. There is only a single sector gap between sectors 35 and 37. 
Therefore, the sector 27--39 data are treated here as
a single light curve. Thus, two year long light curves separated 
by 12 months, with a one sector gap in the second one, are available, followed
by a third $\approx$220~d light curve after an interval of another 18 months.

The TESS light curves of AH~Pic (Fig.~\ref{ahpic-lc} (left))
are characterized by the continuous series of low scale outbursts which first
drew my attention to the AH~Pic phenomenon. A simple visual inspection
shows that whereas the outburst amplitude remains reasonably constant the
intervals between outbursts and the light curve profile over the outburst
cycle exhibits some variability, however, not in a random fashion.
They evolve gradually on time scales longer than the outburst cycle.

The continuous series of outbursts was observed over an even longer time
base by the ASAS-SN. The light curve, spanning a total of ten observing
seasons, and which overlaps with the TESS light curves, is shown in 
Fig.~\ref{ahpic-lc} (right). The upper panel contains the entire light 
curve. The individual outbursts are resolved in the lower frames where
the light curve is shown in expanded form, covering one observing season
in each frame. $V$ magnitudes are plotted in blue, $g$
magnitudes in red. The outbursts continue over the entire time interval
covered by the light curve, but it is obvious that there are periods
with longer and shorter outburst intervals. Moreover, in particular during
the last observing season the minima between the outbursts sometimes 
attain a significantly fainter magnitude than otherwise (I draw
attention to the expanded magnitude scale in the bottom frame of
Fig.~\ref{ahpic-lc}). Such 
dips also observed in CM~Del, KIC~9202990 and in particular in
PS~Eri (see below).

\subsection{V1116 Cephei}
\label{V1116 Cep}

\object{V1116 Cep} (also known as \object{HS 0229+8016}) was identified
by \citet{Aungwerojwit05} as a cataclysmic variable among the stars of the
Hamburg Quasar Survey. They measured a spectroscopic 
orbital period of 0.16149(3)~h. Photometric variations compatible with this
period were also detected but did not permit to obtain a more precise value.

\begin{figure*}
\figurenum{2}
\label{hs0229-lc}
\plotone{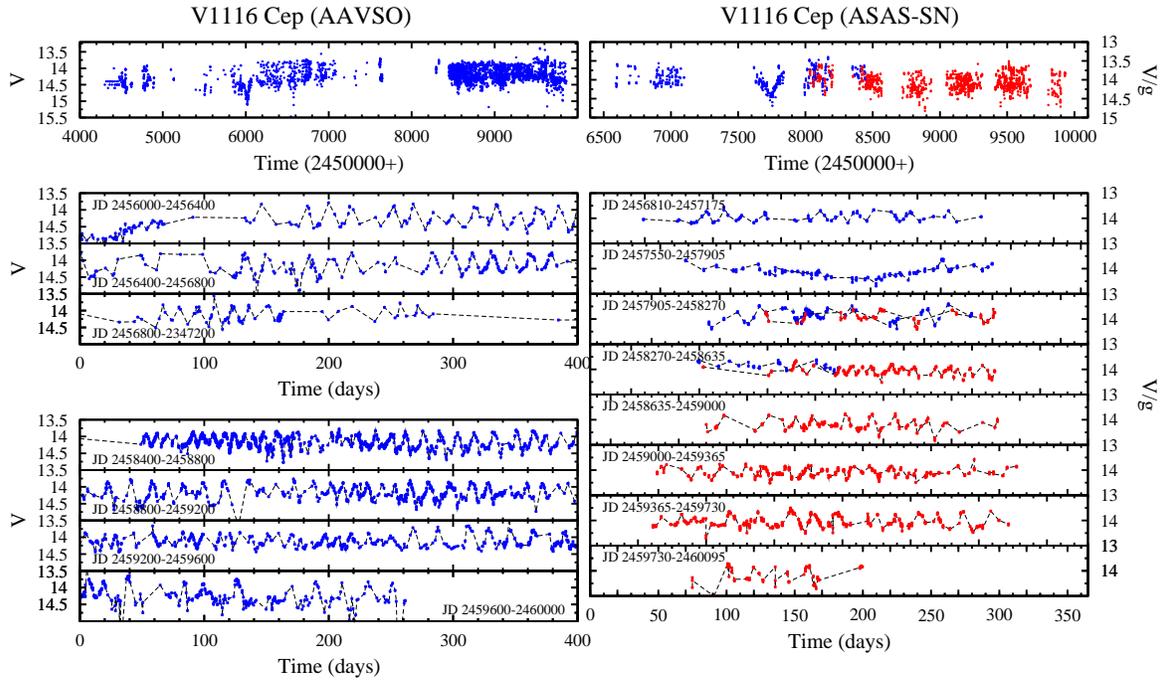}
\caption{AAVSO and ASAS-SN light curves of V1116~Cep, structured in
                 the same way as Fig.~\ref{ahpic-lc}.}
\end{figure*}

Here, I analyse AAVSO, ASAS-SN (Fig.~\ref{hs0229-lc}) and TESS 
(Fig.~\ref{tess-lc}) light curves of V1116~Cep. The AAVSO data cover 
well or reasonably well two time intervals. The ASAS-SN data partly overlap 
with the AAVSO light curve, but also cover well a time interval which is  
only sparsely sampled by the latter, complementing thus the AAVSO data. The 
AAVSO light curve has
also independently been analysed by \citet{Shears20}. A further study of
the long-term variations was presented by \citet{Shears23}. The presence 
of the AH~Pic syndrome in V1116~Cep is clearly confirmed. 

Four TESS light curves are available, three of them 
encompassing two sectors, but the last one covers only one sector 
(Fig.~\ref{tess-lc})\footnote{Based on the quite different aspect of
the first and second TESS light curve, \citet{Shears23} suspect that a 
significant change of the behavior of V1116~Cep has occurred in the
interval beween them (see their Fig.~3). However, this is not the case. No
significant differences are seen in Fig.~\ref{tess-lc}. The apparent change 
of behavior mentioned by \citet{Shears23} can be explained by their use of
PCDSAP data which contain serious distortions in particular in the second 
light curve (see Sect.~\ref{The data}).}. 
All of them contain several brightenings. Although these TESS data are not
as extensive as those of AH~Pic, they also confirm the continuous
outburst sequence characterizing the AH~Pic stars.

I note two special features in V1116~Cyg (see Fig.~\ref{hs0229-lc}). 
(1) An absence of the outburst activity occurred during a dip in the light 
curve between late 2016 and early 2017 ($\approx$JD~2457750). A similar episode 
may have occurred in 2012 (around JD 2456000). (2) The range of 
outburst intervals is much wider than in other systems. Compare, for instance, 
the first and the fourth panel from top of the high resolution AAVSO light 
curve.

\subsection{CM Delphini}
\label{CM Del}

\object{CM~Del} was discovered as a CV by \citet{Bond78}. It remains a poorly
studied object. \citet{Szkody85} presented an optical spectrum, and 
\citet{Lyons01} investigated the UV spectrum. \citet{Shafter85}
measured a spectroscopic orbital period of 0.162~d. But this is based on
few and ill-distributed data points. It differs substantially from the true 
period of 0.14152(3)~d \citep{Bruch24b}. \citet{Zsidi23}
noted the presence of continuous small outbursts in the long-term light
curve of CM~Del, but they did not draw any inferences from this observation.

\begin{figure*}
\figurenum{3}
\label{tess-lc}
\plotone{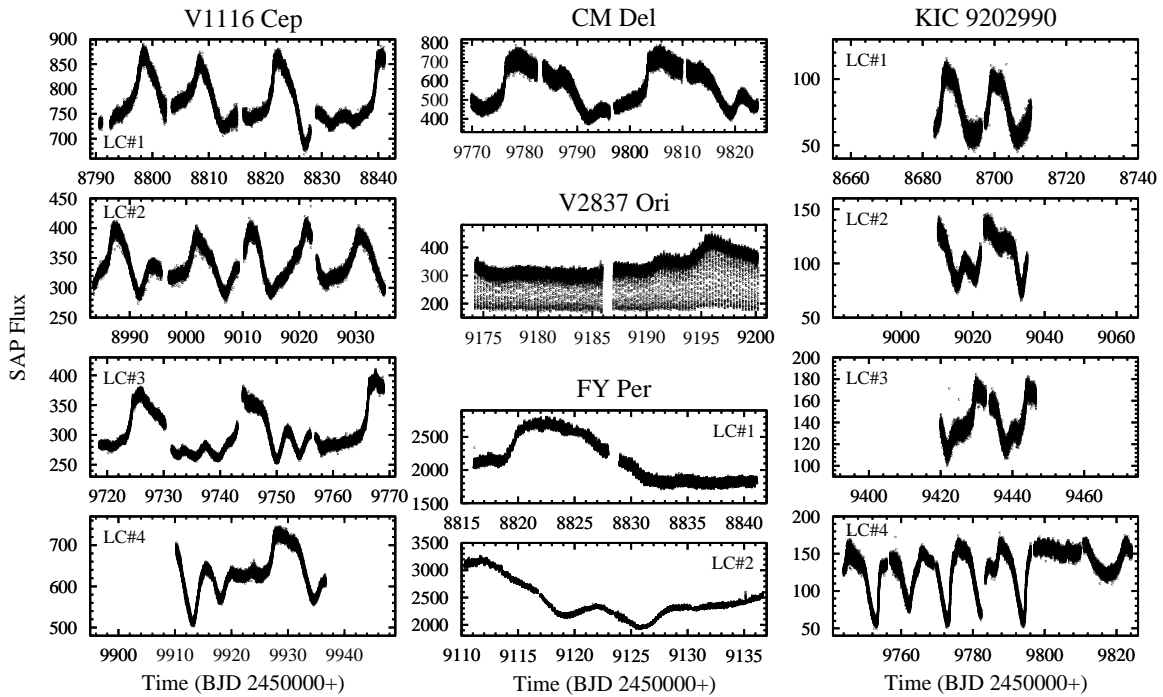}
\caption{TESS light curves of V1116~Cep, CM~Del, V2837 Ori, FY~Per,
                 and KIC~9202990.}
\end{figure*}

\begin{figure}
\figurenum{4}
\label{cmdel-lc}
\plotone{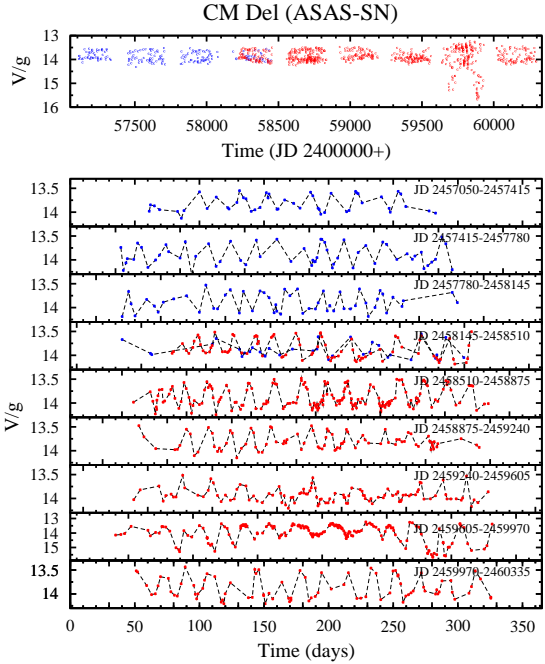}
\caption{ASAS-SN light curve of CM~Del, structured in
                 the same way as the right side of Fig.~\ref{ahpic-lc}.}
\end{figure}

TESS observed CM~Del in two consecutive sectors. The combined light
curve (Fig.~\ref{tess-lc}) contains two brightenings of almost identical
shape. This is not sufficient to identify the AH~Pic syndrome in CM~Del
which is, however, impressively 
demonstrated in the ASAS-SN long-term light curve (Fig.~\ref{cmdel-lc})
consisting of a continuous and uninterrupted series of stunted outbursts 
over the nine years covered by the observations. Whereas their amplitude
is in general fairly constant, two intervals in 2022 are noteworthy when
CM~Del drops to significantly lower magnitudes between the outbursts (note
the enlarged vertical scale in the second panel from the bottom in 
Fig.~\ref{cmdel-lc}). 

\subsection{PS Eridani}
\label{PS Eri}

\begin{figure}
\figurenum{5}
\label{j0229-3958-pow}
\plotone{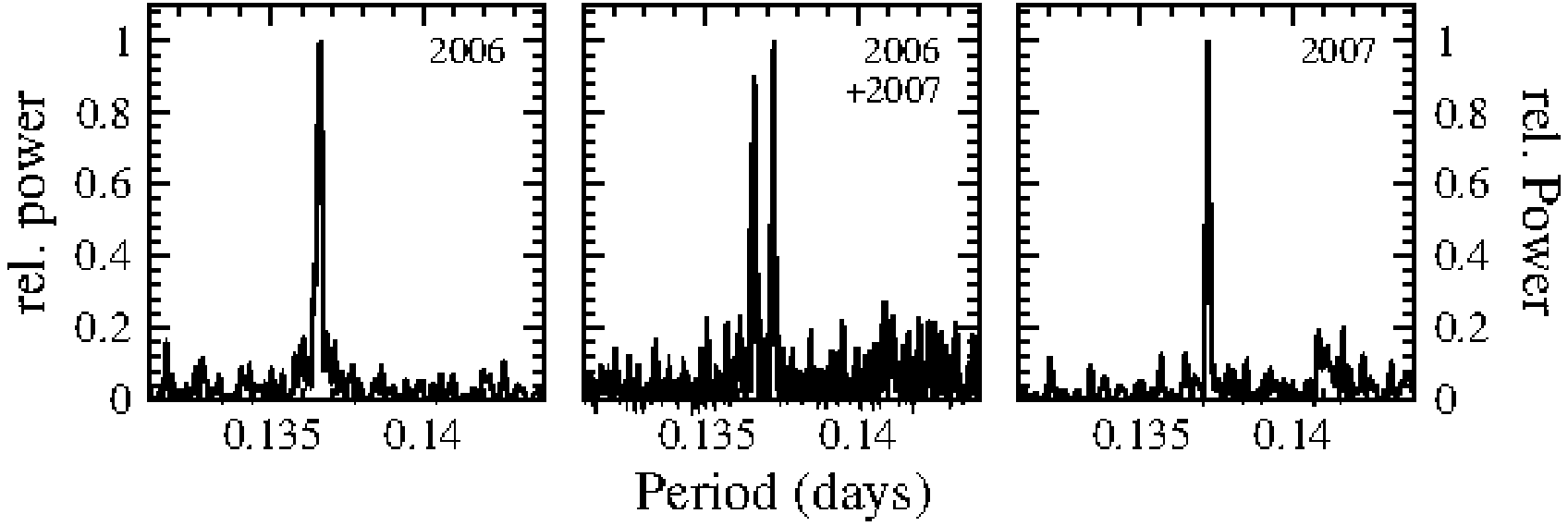}
\caption{Power spectra of the SWASP light curve of PS~Eri of 
the (from left to right) the 2006, the 2007+2007, and the 2007 observing 
seaons in a narrow range around a short term period possibly caused by
superhumps with a slightly different frequency in the two seasons.} 
\end{figure}

The ROSAT X-ray source 1RXS~J022917.1-395851 (= PS~Eri)
\citep{Voges00} is a poorly
investigated object. The only dedicated study was published by
\citet{Hummerich14} who classified the system as a VY~Scl-type novalike
variable. Based on SWASP data \citep{Drake09} they found a period of
0.136621~d in the first part (2006 observing season) of this light curve, 
and in 2007 after a gap of 240~d,
the simultaneous presence of the same period plus a period of 0.137206~d. 
They speculate that the two periods may be due to orbital modulations plus
a superhump. However, their findings are wrong. The two parts of the SWASP
light curve have approximately the same length. Thus, their power spectra
should have a similar resolution. But the right panel of Fig.~3 of
\citet{Hummerich14}, supposed to contain the power spectrum of light curve
part 2 and showing both periods, has a much higher resolution 
than the left panel (part 1). This suggests that the right panel is based on
the combined data, not only part 2. To confirm this suspicion I used the
same SWASP data. The three panels of Fig.\ref{j0229-3958-pow} show the power 
spectra of part 1 (2006, left), part 1+2 (2006+2007, middle) and part 2 
(2007, right). It is
obvious that part 1 only contains the 0.136621~d and part 2 only the 0.137206~d
period. Assuming one of them to be orbital and the other one due to a 
superhumps leads to a superhump excess of 0.0043, much too small for either
a positive or a negative superhump \citep{Bruch23b}. However, considering
that the exact periods of superhumps in novalike variables are known to be 
slightly variable, and that they may be active even if no orbital variations
are seen \citep{Bruch19, Bruch23a, Bruch23b}, both signals may be caused by 
superhumps (either negative or positive). 
The VY~Scl classification of PS~Eri is mainly 
sustained by a deep minimum seen in Siding Spring Survey data 
\citep[see Fig.~1 of][]{Hummerich14}.

\begin{figure}
\figurenum{6}
\label{j0229-3958-lc}
\plotone{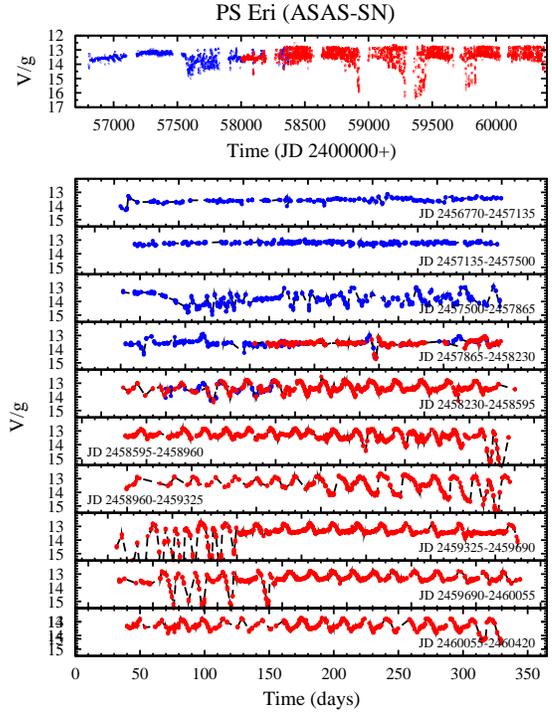}
\caption{ASAS-SN light curve of PS~Eri, structured in
                 the same way as the right side of Fig.~\ref{ahpic-lc}.
                 In the lower frames some deep dips in the light curve
                 are truncated.}
\end{figure}

The relationship of PS~Eri to the AH~Pic stars is already
hinted at in the SWASP data of 2007, but the sequence of stunted outbursts 
is not clearly seen in later years \citep[Figs.~1 and 2 of][]{Hummerich14}.
The ASAS-SN light curve (Fig.~\ref{j0229-3958-lc}) shows that only during
the 2018, July, the continuous train of stunted outbursts defining the
AH~Pic syndrome starts and continues until the end of the light curve.
It is striking that more often than other AH~Pic stars PS~Eri
exhibits deep dips between outbursts. Moreover, even before the AH~Pic 
behavior sets in the system undergoes rather strong but irregular variations
(JD~2457560--2457830), while it is quite constant in the preceeding observing
season. 
 
\subsection{V2837 Orionis}
\label{V2837 Orionis}

Discovered as a variable star a long time ago by \citet{Hoffmeister63}, 
\object{V2837 Ori} (= NSV~1907) remained largely 
unstudied until recently, when \citet{Hummerich17} identified it as a deeply 
eclipsing novalike cataclysmic variable, possibly of the SW~Sex subclass. 
The eclipses permit to measure an accurate orbital period of 0.27610702(2)~d
\citep{Bruch24a}. Thus, V2937~Ori is on the longer side of the CV period 
distribution. 
\citet{Hummerich17} also observed a 4.2~d modulation which they interpret
as the nodal precession period of an accretion disk and which they used to
predict a period of 0.2591~d for a (negative) superhump. An analysis of the
TESS data by \citet{Bruch23a} confirmed the presence of superhumps.

\begin{figure*}
\figurenum{7}
\label{v2837ori-lc}
\plotone{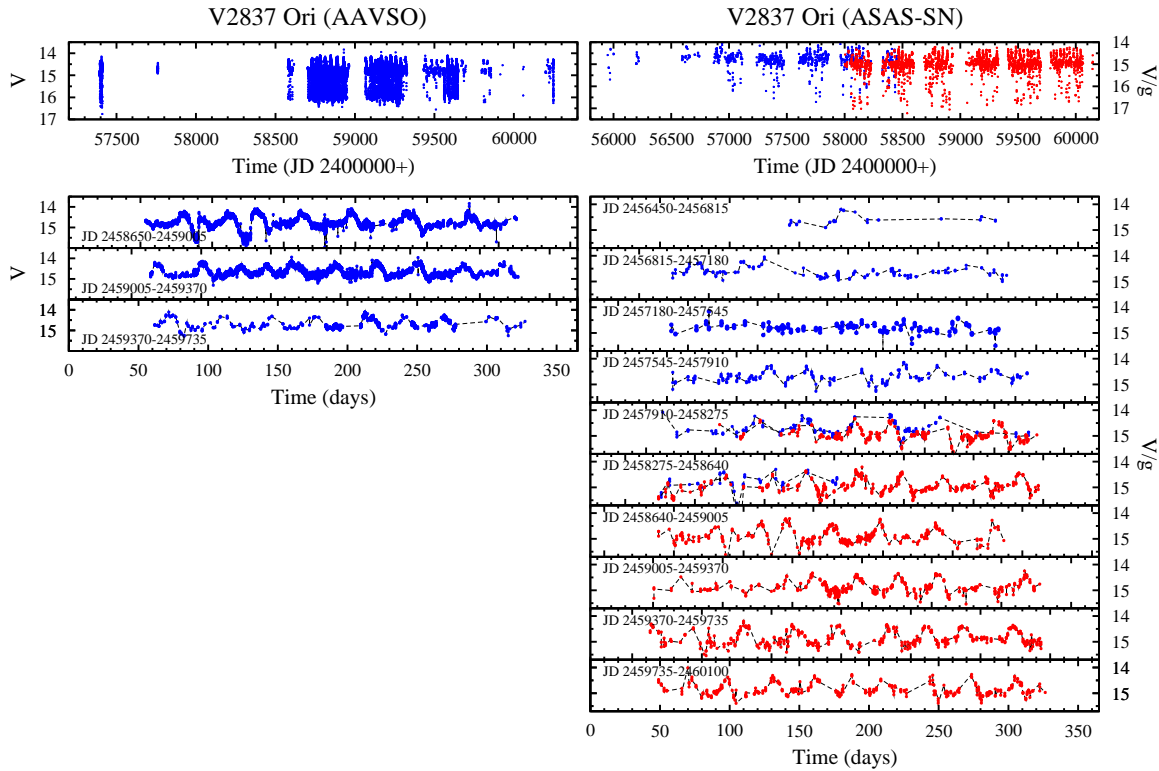}
\caption{AAVSO and ASAS-SN light curves of V2837~Ori, structured in
                 the same way as Fig.~\ref{ahpic-lc}. In the high
                 time resolution versions of the light curve the eclipses
                 have been removed.}
\end{figure*}

The single sector TESS light curve of V2837~Ori (Fig.~\ref{tess-lc})
only contains a single brightening and possibly the end of the decline
from another outburst right at the beginning. This can only suggest the
presence of the AH~Pic syndrome that is, however, confirmed by the AAVSO 
and ASAS-SN light curves. Structured in the same way as 
Fig.~\ref{ahpic-lc} they are shown in Fig.~\ref{v2837ori-lc}. 
In the high resolution version, data points in a phase range of $\pm$0.1 
around the eclipse minima as calculated using the ephemeris of 
\citet{Bruch24a} were removed in order to better visualize the continuous 
outburst activity. The densely covered part of the AAVSO light curve in the 
2019 -- 2021 observing seasons clearly shows the continuous train of small
outbursts which defines the AH~Pic syndrome. However, the longer time base 
of the ASAS-SN data reveals that the latter only started in late 2016, 
being absent in the three preceding observing seasons. It then persists to 
the end of the available data in 2023. Thus, the AH~Pic type behavior is 
not permanent in V2837~Ori. 

\subsection{FY Persei}
\label{FY Persei}

\object{FY Per} is a poorly studied object. No time resolved spectroscopy has 
been published, and photometric studies are limited to its long-term behavior. 
In fact, even its CV nature was questioned \citep[see][for a short 
discussion]{Downes95}, but this issue appears to have been resolved
\citep[see, e.g.,][]{Honeycutt01}. \citet{Bruch23b} detected negative superhumps
in TESS light curves of FY~Per. At 0.25837~d, he also measured the most 
precise orbital period.  

\begin{figure*}
\figurenum{8}
\label{fyper-lc}
\plotone{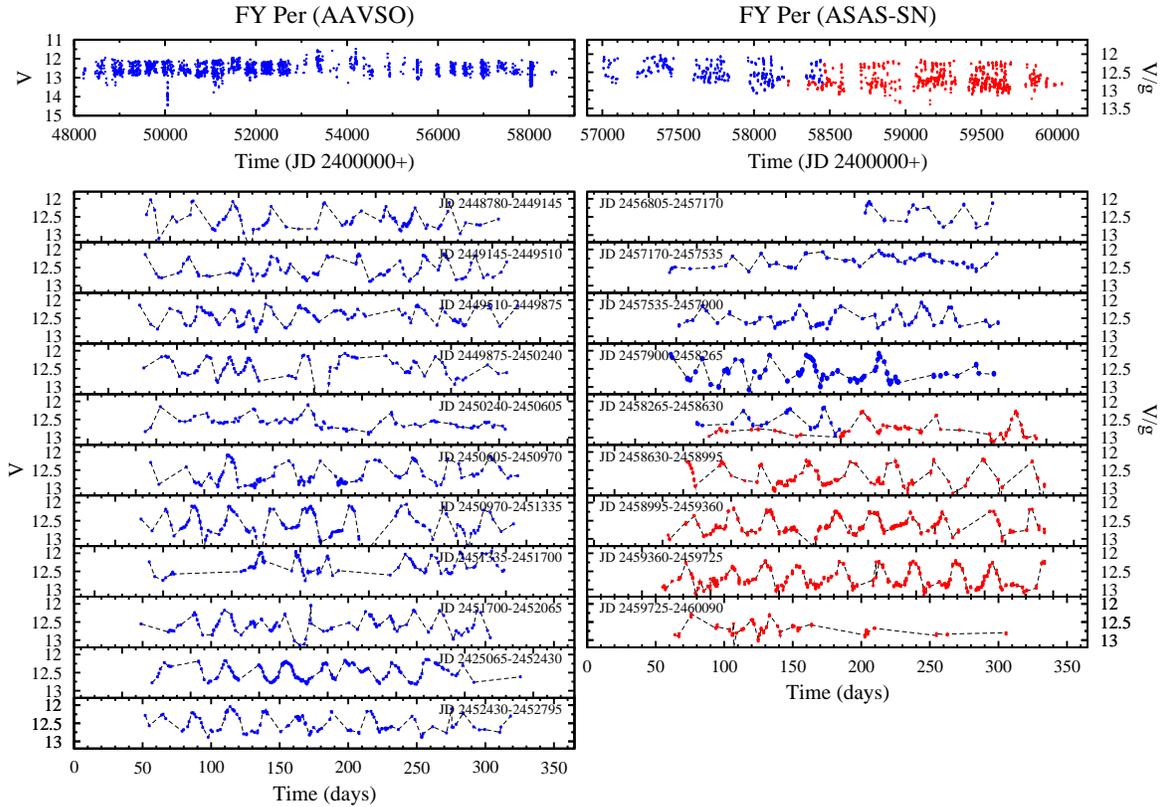}
\caption{AAVSO and ASAS-SN light curves of FY~Per, structured in
                 the same way as Fig.~\ref{ahpic-lc}.}
\end{figure*}

Two TESS light curves of FY~Per are available. Each one encompasses only a 
single sector. Both light curves (Fig.~\ref{tess-lc}) contain a complete or a
partial brightening. Stunted outbursts have been observed in FY~Per
before. First mentioned by \citet{Sazonov92}, many more were seen by 
\citet{Honeycutt01}. \citet{Honeycutt04} mention the `common occurrence
of 0.6~mag oscillations with a characteristic interval of 20--25~days'.
The continuous presence of stunted outburst and thus the AH~Pic syndrome 
is beautifully confirmed by the extensive data sets of ASAS-SN and AAVSO 
(Fig.~\ref{fyper-lc}). The AAVSO light curve covers the
time interval between 1990 and 2019, but only between 1992 and 2003 it
has a dense enough coverage to clearly reveal the successive train of
outbursts. Complementing these data, the ASAS-SN light curve covers the 
more recent years between 2015 -- 2023. As in AH~Pic there are periods of
longer and shorter outburst intervals, but overall, they are considerably 
longer than in AH~Pic.

\subsection{KIC 9202990}
\label{KIC 9202990}

In a limited amount of early Kepler observations
\citet{Ostensen10} noticed that the light curve of \object{KIC 9202990}
is dominated by long term variations superposed upon regular variations with
a period of 0.166~d. They classified the system as a novalike cataclysmic
variable and interpreted the 0.166~period as orbital. Based on the entire
Kepler data set of KIC~9202990, \citet{Ramsay16} investigated the system
further and concluded that the long-term variations constitute a continuous
train of stunted outburst (i.e.\ the AH~Pic syndrome). Short term (flickering) 
variations in KIC~9202990 were studied by \citet{Bruch22}

\begin{figure*}
\figurenum{9}
\label{kic9202990-lc}
\plotone{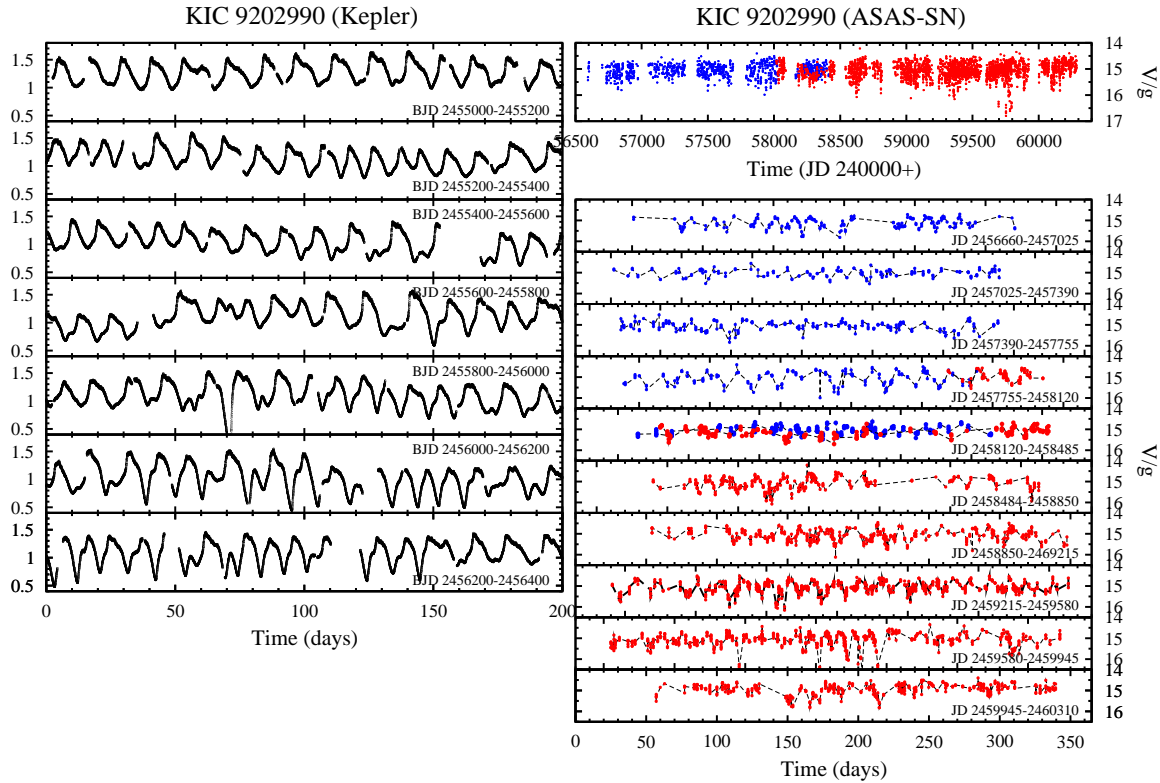}
\caption{Kepler long-cadence light curve of KIC~9202990 (left) and
         its ASAS-SN light curve (right) structured in
                 the same way as the right side of Fig.~\ref{ahpic-lc}.}
\end{figure*}

Kepler observed KIC~9202990 for more than four years almost continuously in
long cadence mode. During three limited time intervals within these four
years short cadence light curves are also available 
\citep[see Fig.~5 of][]{Bruch22}. The full long cadence light curve,
reproduced in Fig.~\ref{kic9202990-lc} (left), is a most impressive example
of the AH~Pic type behavior.

The temporal coverage of KIC~9202990 by TESS is much less extensive. TESS
observed the star on four occasions (Fig.~\ref{tess-lc}), three 
times during a single sector, while the last visit encompassed three sectors.
A peculiarity of the latter light curve is the cessation of the outburst
activity at the end. The existence of intervals without the usual brightenings
is also observed in the ASAS-SN light curve (Fig.~\ref{kic9202990-lc}, right),
which starts after the end of the Kepler observations and extends to the 
present day. It thus nicely complements the close coverage of KIC~9202990 
by Kepler. Finally, just as other AH~Pic systems, sometimes the minimum 
between outbursts attains significantly fainter magnitudes than usual.

%\begin{figure}
%\figurenum{1}
%\label{kic9202990-asassn-lc}
%\plotone{kic9202990-asassn-lc.eps}
%\caption{ASAS-SN light curve of KIC~9202990, structured in
%                 the same way as Fig.~\ref{ahpic-lc}.} 
%\end{figure}

\section{Properties of the AH~Pic stars}
\label{Properties of the AH Pic stars}

\subsection{Outburst intervals and amplitudes}
\label{Outburst intervals and amplitudes}

The long available light curves of the target stars permit the determination
of the outburst intervals (or the outburst recurrence time) and their
amplitudes. To this end, I employ the ASAS-SN and AAVSO data of all
systems, as well as the long cadence Kepler light curve of KIC~9202990
and the TESS light curves of AH~Pic. The TESS data of the other systems
are too sparse for statistically meaningful results.

Given the quite different sampling of the TESS and Kepler data on the one hand, 
and the ASAS-SN and AAVSO data on the other, slightly different approaches 
were adopted. In order to remove orbital variations and random modulations
such as flickering, the TESS and Kepler data were first smoothed with a 
Savitzky-Golay \citep{Savitzky64} filter with a cutoff time scale of 2~d and
a fourth order smoothing polynomial. This reduces the impact of orbital 
variations on the outburst amplitude.
The smoothed light curves were transformed
from the flux to a magnitude scale with an arbitrary zero point. The epochs
of individual maxima and minima of each outburst cycle were then measured. 
The average of the time intervals between subsequent maxima and subsequent 
minima is taken as the outburst recurrence time. Similarly, the amplitude 
was measured as the magnitude difference between a light curve minimum 
to the next outburst maximum, and again from maximum to the next minimum. 
% The accuracy of these amplitudes 
% depends on the accuracy of the flux zero point of the Kepler and TESS
% data. In particular the TESS data could be compromised by contamination 
% from neighbouring stars or an inadequate background subtraction, considering 
% the coarse spatial resolution of TESS. However, I have no indication for a
% significant error of this kind in the case of AH~Pic, given that the average 
% flux in the many different sectors of which the light curve is composed is 
% quite constant, and that the next neighbour to AH~Pic of similar brightness
% is $\approx 45\arcsec$ away.

To measure outburst intervals and amplitudes in the ASAS-SN and AAVSO light
curves, only those parts were used where a continuous train of outbursts 
could clearly be identified. The time resolution of these data is, of course, 
much inferior to that of the TESS and Kepler data. Therefore, a minimum or 
maximum point in the light curve may be somewhat different from the true 
minimum or maximum. This will enhance the scatter of the time interval between 
outburst, but should not affect its mean or median value significantly. 
However, the amplitude will have a systematic error, being smaller than it 
really is. This effect may be somewhat compensated by random effects such
as measurement errors, flickering and irregularly sampled orbital variations
because on average these will push the extrema to more 
extreme values. In comparing the amplitude derived from the ASAS-SN or AAVSO 
to those of the Kepler or TESS light curves the different passbands must 
also be considered. While the terrestrial observations were all taken in the 
$V$ or the (quite similar) Sloan $g$ band, the TESS satellite has a wide 
passband between 6\,000 and 1\,0000~\AA, centered on the Cousins $I$ band. 
Kepler has a similarly broad passband, but offset by roughly 1\,000~\AA\ 
to the blue.

\begin{figure*}
\figurenum{10}
\label{distr}
\plotone{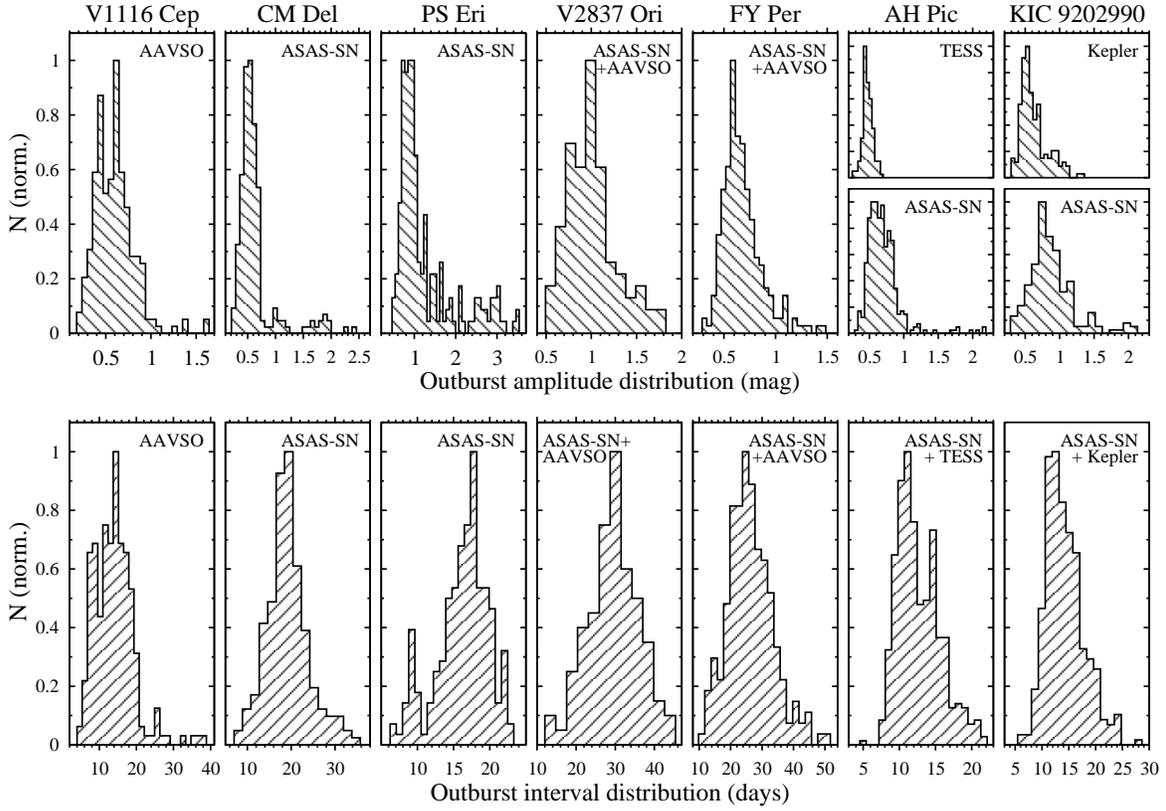}
\caption{Distribution functions of outburst amplitudes and intervals of the
         AH~Pic stars.}
\end{figure*}

The median outburst amplitudes and intervals are listed in 
Table~\ref{Table: AH Pic stars}. I prefer to quote the median instead of
the mean values because of the sometimes skewed distribution of individual
outburst intervals and amplitudes. The distribution functions are shown in
Fig.~\ref{distr}. They extend over a wide range of typically 1.5 times the
median value. In order to parameterize their width the standard deviation
of a Gaussian fitted to the distribution function, divided by the median
value (thus, the relative width), is also listed in 
Table~\ref{Table: AH Pic stars}. Most of the amplitude 
distributions have an extended tail towards high amplitudes. This reflects
the fact that the light curves sometimes drop to an unusually low minimum 
after an outburst. This is seen at the end of the AH~Pic and CM~Del
light curves in Figs.~\ref{ahpic-lc} and \ref{cmdel-lc}, respectively,
and most clearly in the PS~Eri light curve
(Fig.~\ref{j0229-3958-lc}). 
At the longer wavelengths observed by TESS and Kepler, the amplitude is 
significantly smaller than in the visual range (ASAS-SN and AAVSO). This 
can probably be
attributed to the stronger contribution to the total light of the late type 
secondary star star in the red and/or a spectral energy distribution
of the outbursting light source that is bluer than the quiescent light.
Most of the amplitude distributions have a 
single maximum. But in V1116~Cep it appears to be bi-modal. A
bi-modal distribution is also suggested in the outburst intervals 
of AH~Pic and PS~Eri. Otherwise, the overall interval 
distributions are fairly Gaussian in shape. However, as mentioned above,
variations of the outburst interval do not occur randomly but there may be 
periods with systematically longer or shorter intervals.

The ASAS-SN and AAVSO data are not well enough sampled to permit a reliable 
statement on the outburst shape. Regarding the TESS and Kepler light curves
it appears that there is a continuum of shapes even in a given system that
at one extreme consists of a steep rise to maximum, followed by a slowly
declining plateau, and then by a more rapid decline (but less so than the 
rise) ending the outburst. The duration of the plateau is not constant 
but may shrink, until at the
other extreme it disappears altogether and the decline towards minimum starts 
right after the maximum. Just as in the case of the outburst intervals the
transition between these extremes occurs only gradually over several outburst
cyles. Moreover, there appears to be a positive correlation between the length 
of the plateau phase and the outburst interval. 
 
\subsection{Absolute magnitudes}
\label{Absolute magnitudes}

In order to compare the absolute magnitude of the target stars with those
of other novalike variables (and old novae) and with dwarf novae 
the average $V$- and $g$-band magnitudes of the minima between stunted 
outbursts (disconsidering obvious dips below the normal minima) were 
determined in all ASAS-SN light curves. In order to reduce the $g$ band 
magnitudes to the $V$ band, the difference between the average $g$ and $V$ 
values in the overlap region of the two band in the ASAS-SN light curves 
was subtracted from all $g$ band magnitudes\footnote{This did not work for
FY~Per because the overlap is small and no outburst $g$ magnitudes were
measured. Therefore, in this case the difference between $V$ and $g$ averaged
over the entire light curve was used.}. Distance measurements based on
Gaia DR3 data were taken from \citet{Bailer-Jones21}. Together with the
interstellar absorption calculated using the reddening $E_{B-V}$ estimated 
in the tri-dimensional Stilims extinction maps 
\citep{Lallement14, Capitanio17} these yield the abolute visual magnitude 
$M_V$ of the AH~Pic stars as listed in Table~1.
% Table-\ref{Table: AH Pic stars}
Distribution functions of the absolute $V$ band magnitudes of novalike 
variables (including also old novae in the sample) and dwarf novae during 
quiescence and outburst maximum were determined as detailed in 
Appendix~\ref{Absolute magnitude distribution functions}. Only SS~Cyg- and
Z~Cam-type dwarf novae were considered, avoiding the on average much
fainter short period SU~UMa- and WZ~Sge-type stars. 

\begin{figure}
\figurenum{11}
\label{Mv-distr}
\plotone{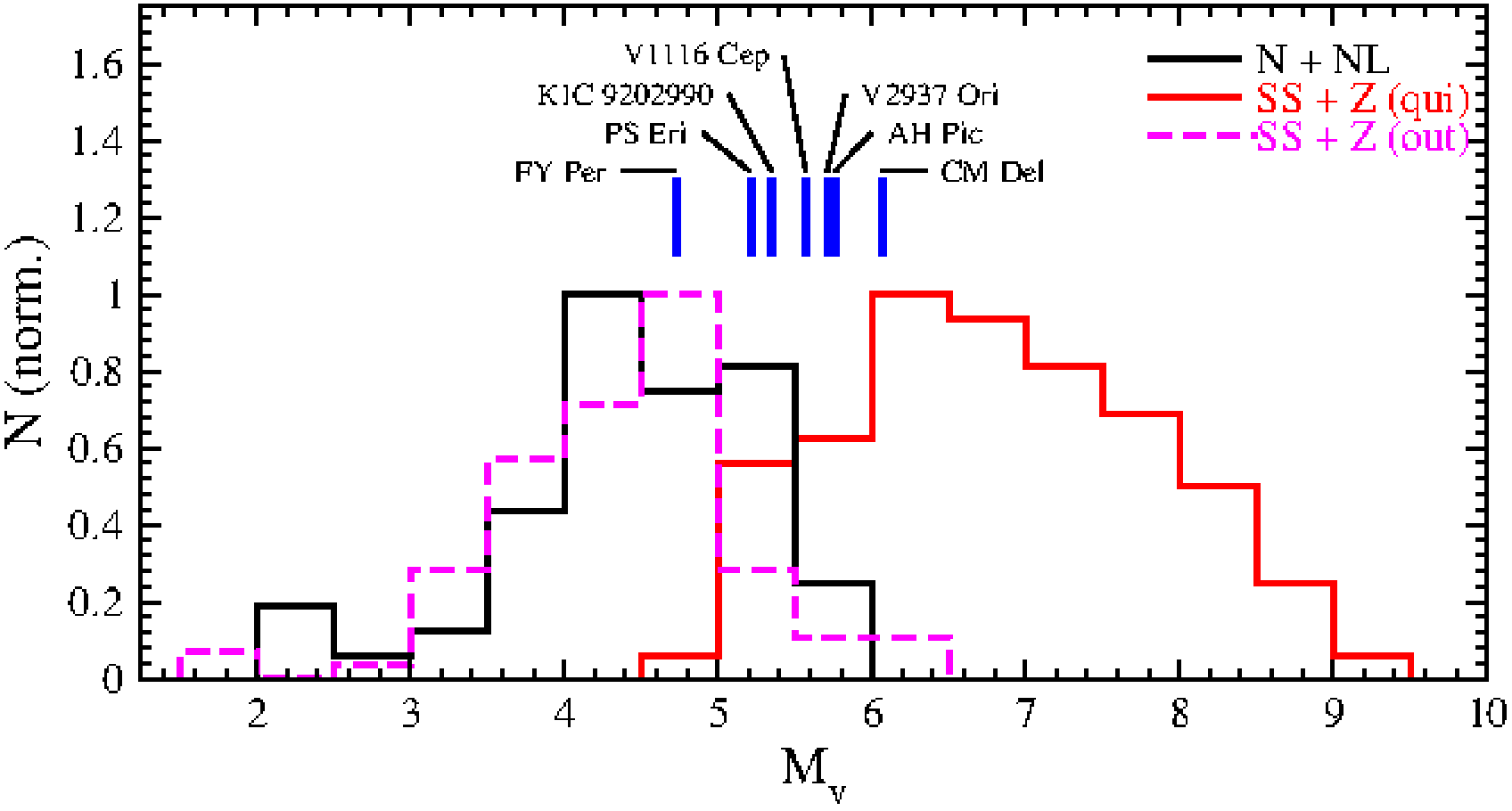}
\caption{Distribution functions of the absolute $V$ band magnitude of
novalike variables and old novae (solid black), and of SS~Cyg- and Z~Cam-type 
dwarf novae during quiescence (solid red) and outburst maximum (dashed 
magenta). The blue bars indicate the absolute magnitudes of the AH~Pic stars.}
\end{figure}

The distribution functions are show in Fig.~\ref{Mv-distr}. The absolute
magnitudes of the AH~Pic stars are indicated as blue bars. They are all
fainter than the average $M_V = 4.42$ of novalike stars,
and brighter than the average $M_V = 6.85$ of dwarf novae, lying
in the overlap region between the two distribution functions.

\subsection{Orbital variations}
\label{Orbital variations}

The high time resolution of the TESS and Kepler data, available for all
target stars except PS~Eri, permits to study their
variability on orbital time scales. As will be seen,
the AH~Pic stars share some common properties also concerning their variations
around the binary orbit. The orbital periods of all the stars are listed in 
Table~\ref{Table: AH Pic stars}. The improved value for FY~Per is based on
the frequency of the power spectrum peak of TESS LC\#2 shown in Fig.~15 of
\citet{Bruch23b}.

The accurately known orbital periods permit to construct high quality 
orbital waveforms. To this end, the pre-whitened light curves (that is, after 
subtraction of its smoothed version; see
Sect.~\ref{Outburst intervals and amplitudes}) 
were folded on the orbital period. To 
reduce the noise the waveforms were binned into intervals of width 0.01 in 
phase. This was done for each TESS light curve and, in the case of KIC~9202990, 
for the short cadence Kepler light curves\footnote{The long cadence data 
yielded an identical but noisier orbital waveform and are therefore not 
further considered in this context.}. As mentioned by \citet{Bruch23b}, instead 
of orbital variations only a strong negative superhump is visible in 
TESS light curve LC\#1 of FY~Per. In contrast, LC\#2 
exhibits orbital variations along with superhumps. Therefore, only the 
latter is suited to study the orbital waveform of FY~Per. 

\begin{figure*}
\figurenum{12}
\label{waveforms}
\plotone{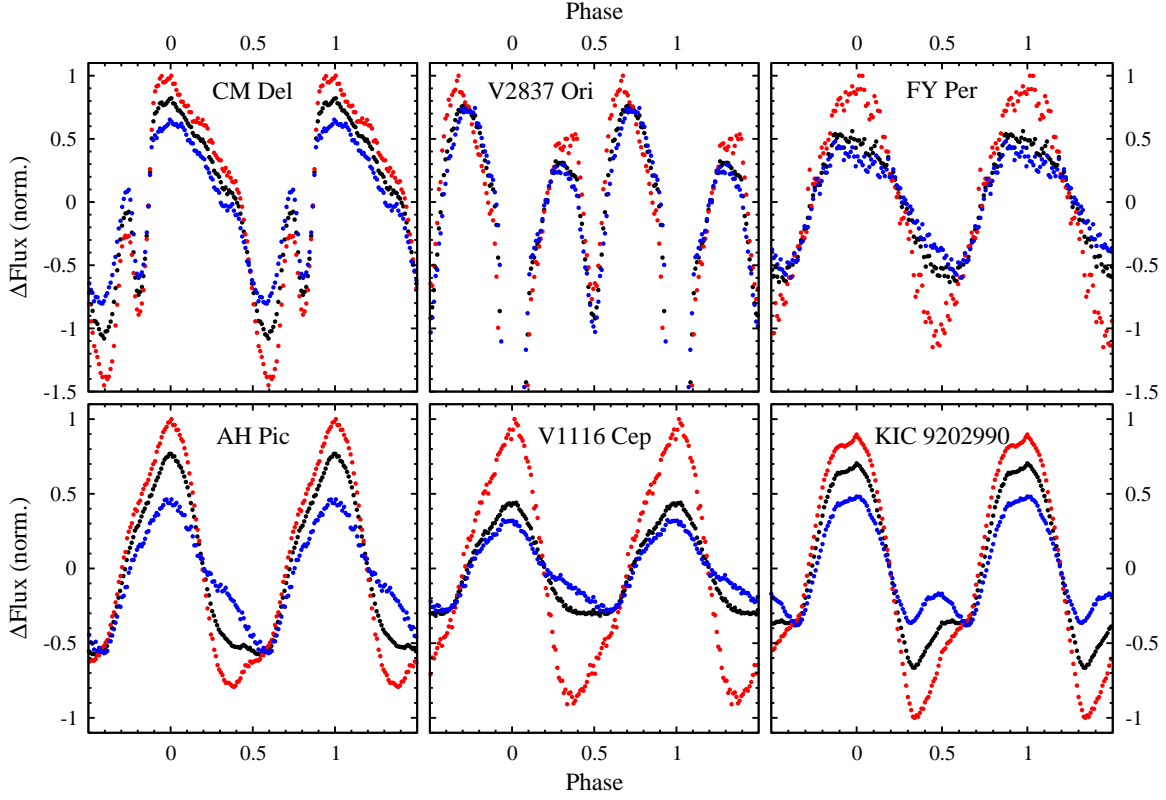}
\caption{Orbital waveforms of the AH~Pic stars. The black
                graphs refers to the entire light curves, the red ones to
                outburst, and the blue ones to inter-outburst phases.} 
\end{figure*}

In all systems with more than one light curve the individual data sets yielded
identical results. Thus, the waveforms are stable over time and the
average for these stars was constructed, weighting each light curve 
according to the number of contributing data points. The results are shown in
Fig.~\ref{waveforms} as black graphs. The zero point of phase is defined
as the maximum of the prominent hump in the waveforms of all targets,
except in the case of the eclipsing system V2837~Ori, where the center
of the eclipse is taken as phase zero. 

\begin{figure}
\figurenum{13}
\label{ahpic-orb}
\plotone{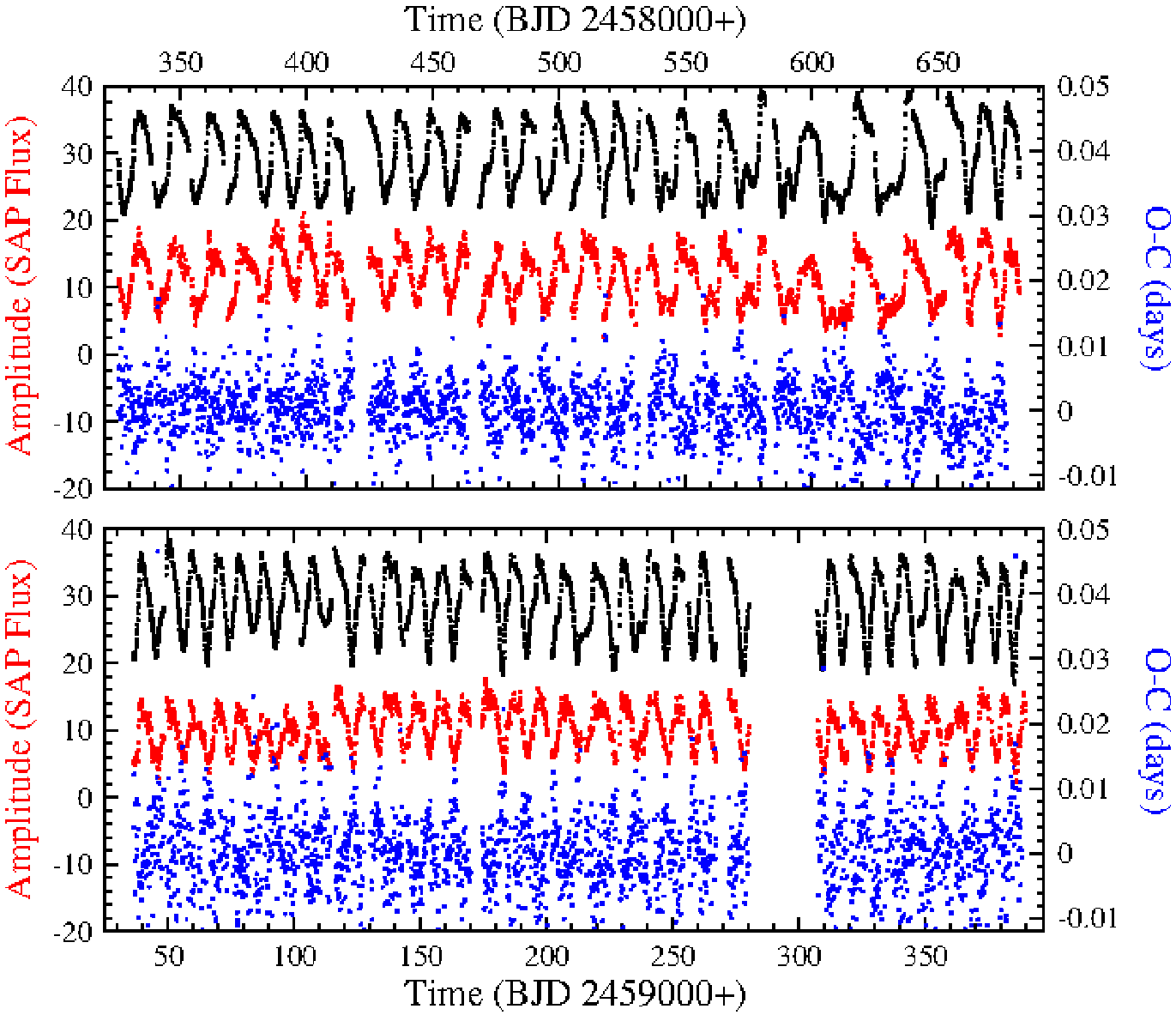}
\caption{Dependence of the orbital variations of AH~Pic on the outburst
phase in light curves LC\#1 (top) and LC\#2 (bottom) of AH~Pic. 
The black graphs contain
the light curves, binned in time intervals of 0.1~d. The red graphs (left
hand scale) show the amplitude of the orbital variations, the blue
one the $O-C$ values (right hand scale) between the observed and calculated
epochs of orbital minima.} 
\end{figure}

However, the average waveforms over the entire light curves are misleading.
Significant details depend on the outburst phase. This is shown in
Fig.~\ref{ahpic-orb} at the example of the two long TESS light curves of
AH~Pic displayed as black graphs in the two frames of the figure,
binned in time intervals of 0.1~d. Beneath, the amplitude
$A$ of the orbital variations are shown in red (left hand scale of the figure).
For each cycle, $A$ was measured as the full amplitude of a sine fit to a
section of the light curve including said cycle plus the preceding and 
following cycles. An alternative approach, fitting parabolae to the minima and 
the maxima of each cycle and determining the flux difference of their extrema
yielded similar but noisier results. A strict correlation is observed between
$A$ and the brightness of AH~Pic. During outburst maximum the amplitude of
the orbital variations is consistently and significantly higher than during 
minima. It increases by a factor of 1.8 during outburst. That the
amplitude of the orbital variations increases during outbursts may not
be surprising, but it is also not entirely trivial. It means that the extra
light during the brightenings is not just added to the inter-outburst light, 
but that it is also modulated around the orbit. The median outburst amplitude
of AH~Pic of 0.45~mag leads to a flux ratio $F_{\rm max}/F_{\rm min} = 1.5$
between outburst maximum and minimum. This is less than the observed
amplitude ratio of the orbital variations. Thus, the extra light during 
outburst is even stronger modulated than the quiescent light.

The blue dots in Fig.~\ref{ahpic-orb} show the difference
$O-C$ between the observed epochs of the orbital minima, given as the time of
minimum of the quadratic fit to the minima of each cycle, and the calculated 
epoch using the orbital period. Again, there is a clear dependence of $O-C$ 
on the outburst phase of AH~Pic. This means that the orbital minimum is 
shifted to earlier phases during outburst. 

In order to visualize the outburst phase dependent differences for all target 
stars. waveforms were 
constructed separately for outburst and inter-outbursts intervals. Since
there is no clear distinction between these stages, they are here taken to be
roughly as the upper and lower parts of the continous outburst trains. For
simplicity, I refer to the latter as quiescence, being
aware that truly quiescent intervals cannot well be defined in AH~Pic
stars. The resulting waveforms are shown as red (outburst) and blue 
(quiescence) graphs, respectively, in Fig.~\ref{waveforms} and are remarkably 
similar for the different objects. 

In the prototype AH~Pic the waveform is dominated by a single hump extending 
over $\approx$70\% of the orbit. In quiescence as well as in outburst it has 
a small kink at phase $\approx$0.8 after which the gradient is slightly 
smaller. During quiescence, halfway down the orbital hump the slope of the 
waveform decreases suddenly and the flux continues to drop linearly until the 
onset of the next hump at phase 0.60. In contrast, during outburst the 
decline from the hump occurs without a break until the minimum at phase 0.37. 
Thereafter a somewhat structured rise, which in the first part resembles the 
egress from a shallow eclipse, is observed until phase 0.60 (the same phase at 
which the linear decline during quiescence gives way to the onset of the hump),
and then the hump starts. 

In almost all details this behavior is repeated in V1116~Cep. The only
difference is the lack of an indication for an eclipse during outburst.
The waveforms of KIC~9202990 were already shown and discussed by 
\citet{Bruch21}. Again, the same pattern is observed, but the eclipse-like
feature is now also present during quiescence. The difference of its 
minimum phase during quiescence and outburst is noteworthy. Moreover, the kink
during the rising phase of the hump now occurs much closer to the hump
maximum. During the brightening an additional spike develops upon the hump.

In both, outburst and quiescence, the CM~Del waveform contains two well
expressed minima of different width which gives the impression of the 
occultation of two light sources of different extension. 
Notwithstanding the somewhat different shape of the orbital
hump, the quiescent waveform resembles very much that of KIC~9202990 during
quiescence. Moreover, just as in AH~Pic and V1116~Cep there is a transition
to a shallower slope (shortly before dropping into the first minimum) on the 
declining branch of the hump which is missing during outburst. The outburst
waveform differs from that of KIC~9202990 in the sense that the second, 
sharper minium is absent in the latter. The phase of the second minimum does
not change between outburst and quiescence, but the phase of the first one
is slightly later during outburst compared to quiescence, just as in 
KIC~9202990.

The FY~Per waveforms are noisier because fewer data are available. 
But again, the pattern close to the
hump minimum during both, quiescence and outburst, is the same as in AH~Pic
and V1116~Cep. However, at hump maximum a difference is obvious during
quiescence. Concerning V2837~Ori, the interpretation of the waveform is 
hampered by the deep primary eclipse and also by the clear presence of 
a secondary eclipse. A 
slight difference of the phase of the maximum after the scondary eclipse is 
observed between outburst and quiescence. Before secondary eclipse the 
differences between outburst and quiescence appears to the compatible with the 
structure of the other AH~Pic stars close to the orbital minimum, namely a rise 
during outburst and a slight decline during quiescence.

\begin{figure*}
\figurenum{14}
\label{wf-time-resolved}
\plotone{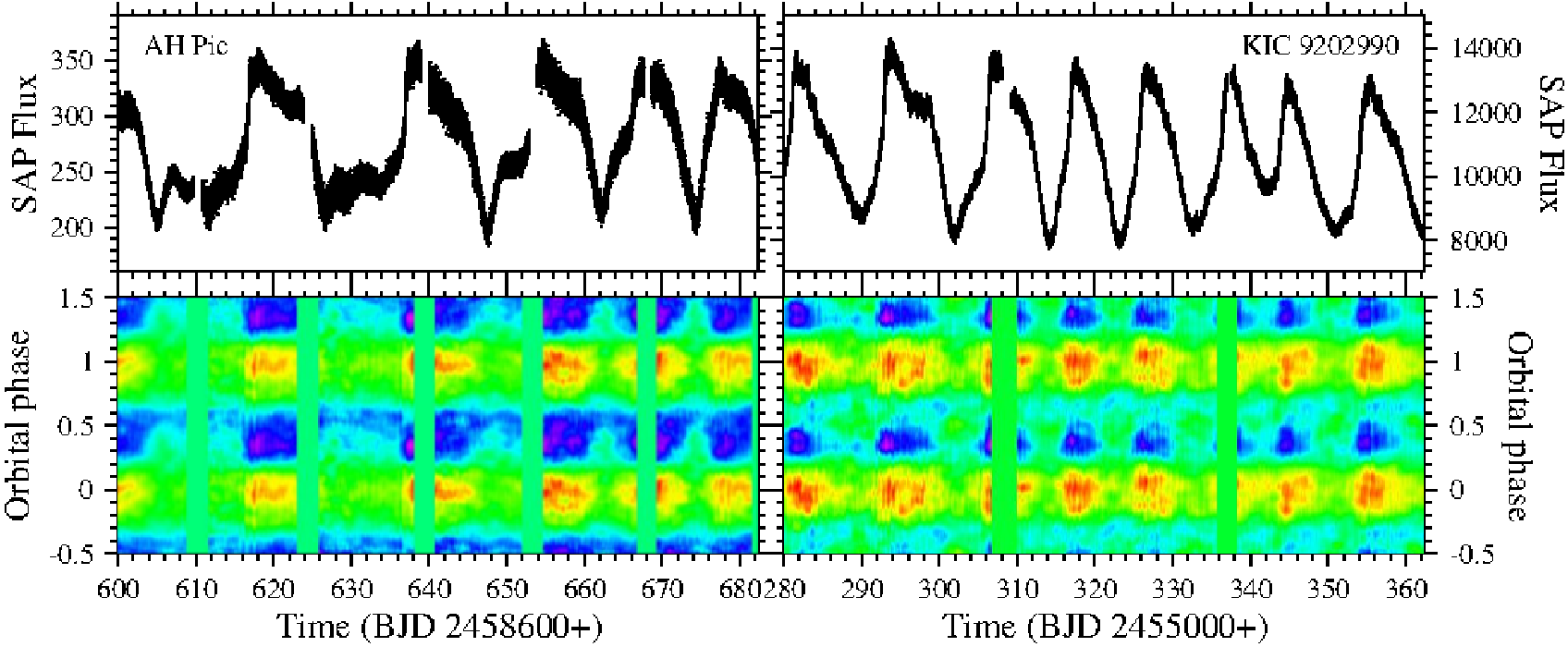}
\caption{{\it Top:} Limited section of the light curves of AH~Pic and
KIC~9202990. {\it Bottom:} Average waveforms in a sliding window with a
width of 10 orbital cycles as a function of time. Orbital minima are 
color-coded in blue/purple, maxima in yellow/red. The vertical green bars
mark intervals during which the sliding window contains less than 10 cylces 
due to gaps in the original light curves and therefore no average waveforms
were calculated.}
\end{figure*}

The consistent change of the waveform over the outburst cycle is not only a 
property of the average over all light curves of a given system but it repeats 
itself in each individual cycle. This is show in 
Fig.~\ref{wf-time-resolved} where in the upper panels a limited interval
of the light curves of AH~Pic and KIC~9202990 is displayed. The lower panels
contain as a function of time the average orbital waveforms in a sliding
window with a width of 10 cycles in a tri-dimensional representation. 
Orbital minima are color-coded in 
blue/purple, maxima in yellow/red. It is obvious that during each outburst
cycle the variations of the waveform are strictly correlated with the
brightness in the same way as in the average waveforms of Fig.~\ref{waveforms}.

Given the considerable variety of orbital waveforms of novalike variables in
general, and their frequent variability on time scales of months and years
\citep{Bruch24a}, their constancy, extreme similarity, and consistent 
variations between outburst and quiescent phases in the AH~Pic stars
is amazing. It points at
a significant likeness of their structure and evolution during the
outburst cycle, emphasizing the kinship of all AH~Pic stars.

\section{Related systems}
\label{Related systems}

\subsection{Other novalike variables with stunted outbursts}
\label{Other novalike variables with stunted outbursts}

\begin{figure*}
\figurenum{15}
\label{synopse}
\plotone{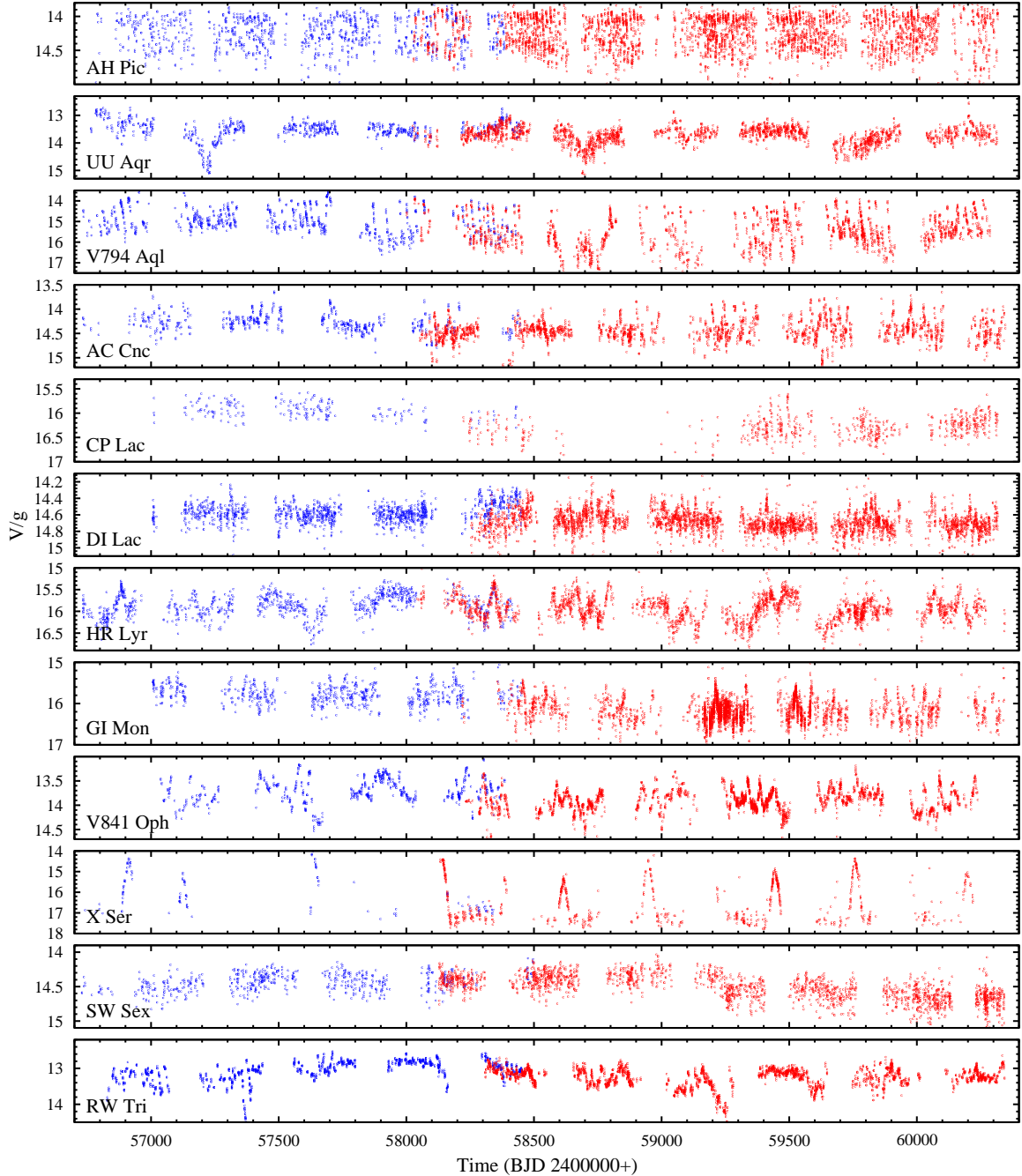}
\caption{Synoptic view of ASAS-SN light curves of NLs with stunted
                 outbursts reported in the literature. As a reference for
                 the aspect of a AH~Pic star light curve on the chosen
                 horizontal (time) scale, the AH~Pic light curve is 
                 reproduced in the upper frame. Blue dots represent data in 
                 the $V$ band; $g$ band data are shown in red.}
\end{figure*}

Stunted outburst in other novalike variables and old novae have been observed
and are characterized in the literature, for instance by
\citet{Honeycutt01} and \citet{Vogt18}. 
In some cases the authors claim that they occur frequently in quick 
succession. This raises the question if these stars could be related to the
AH~Pic stars. Most of the previous studies are based on observations sampled
less frequently and/or over a smaller time base than is available in modern
surveys. Thus, a better coverage may shed light on such a relationship.
For 11 of the systems listed in the mentioned references useful ASAS-SN light 
curves exist. Others -- in particular some old novae observed by 
\citet{Vogt18} -- are too faint for the ASAS-SN. 
The 11 light curves are reproduced in Fig.~\ref{synopse} on the
same time scale. Eclipses in the eclipsing NLs UU~Aqr,
AC~Cnc, SW~Sex, and RW~Tri have been removed. The uppermost frame of the figure 
contains the ASAS-SN light curve of AH~Pic as a reference for the visual 
impression of a light curve with continuous stunted outbursts on the chosen 
horizontal scale. In all cases I have also inspected plots of the seasonal 
light curves on an expanded scale in order to better identify small scale 
outbursts.

Many of the light curves exhibit consistent brightness variations with 
amplitudes of
the order of some tenths of a magnitude on different times scales. This 
makes it sometime difficult to draw a line between brightenings which deserve
the name stunted outbursts and other variations. No outbursts were detected
in SW~Sex. With some confidence, stunted outbursts were occasionally seen in
UU~Aqr, CP~Lac, DI~Lac, HR~Lyr, V841~Oph and RW~Tri. But even if they 
sometimes occur close in time, they remain far from forming the long
sequences of brightenings in immediate sucession that defines the AH~Pic 
syndrome.

AC~Cnc and GI~Mon show some resemblence to the AH~Pic stars. Starting in
the second half of 2019 AC~Cnc exhibits semi-regular brightenings with
amplitudes typical for stunted outbursts. But their duration and the intervals
between them are somewhat longer than in the genuine AH~Pic stars. Moreover,
the brightenings are normally separated by a quiecent interval of appreciable
length. The light curve of GI~Mon contains sections of (noisy) sequences
of variations with an amplitude of roughly 0.8~mag which repeat about very
50 days. There may be thus a kinship between these two systems and the 
AH~Pic star.

X~Ser is a special case. \citet{Honeycutt98a} observed three small scale 
outbursts in the old nova \object{X~Ser} (1903). However, the ASAS-SN light 
curve tells a quite different story. It is punctuated by fairly regular 
outbursts on a much larger scale than normal stunted outbursts. Using also 
AAVSO data, disregarding three outburst intervals longer than 300~d, and 
assuming that some outbursts have been missed, I derive an average interval 
between the brightenings of $241 \pm 18$~d, an average amplitude of 
$2.6 \pm 0.3$~mag and a mean total duration of $62 \pm 5$~d. The amplitude 
is quite normal for a dwarf nova. The comparatively long outburst interval 
and duration is not unusual at the orbital period of $\approx$1.5~d 
\citep{Schaefer22} \citep[see the discussion in][]{Bruch17}. Thus, a 
century after its eruption as a nova X~Ser appears to have transformed 
into an ordinary dwarf nova. For a more detailed analysis of the long-term 
light curve of this object, see \citet{Simon18}.

Of all the stars of this section \object{V794 Aql}
resembles the AH~Pic stars most. \citet{Honeycutt14} studied the long-term
variations between 1990 and 2012 in detail, rectifying some views
expressed in earlier publications. They detected stunted outbursts
`which appear, at least sometimes, as continuous oscillations'. The
ASAS-SN light curves confirms this notion. Only in the 2019 observing
season (JD~2458560-2458820) the continuous outbursts are absent. V794~Aql
is a well-known VY~Scl star and went through a low state during that
year. But even continuing in this low state for at least another 1.5~years
the outburst activity was resumed in the following season. A notable
difference between V794~Aql and the AH~Pic stars is the much larger
outburst amplitude. It is almost always well above 1~mag and can reach
3~mag during the low state. Noting that the brightness at outburst maximum
remains rather constant, the larger amplitude during the low states suggests
that the intrinsic outburst properties do not change, but the outburst 
amplitude is attenuated by a stronger background light during the high
state. Unfortunately, it is not possible to verify if the similarities
between V794~Aql and the AH~Pic stars extend to the orbital waveform.
Although a single sector TESS light curve, containing a 1.2~mag outburst
is available, its power spectrum lacks a convincing signal at or close
to the spectroscopic orbital period of 0.1533~d \citep{Honeycutt98b} or any
other indication of a stable periodicity. 

Recently, \citet{Zsidi23} presented long term light curves of further NLs
exhibiting brightenings similar to stunted 
outbursts. Of these, discounting FY~Per and CM Del as already known 
AH~Pic stars and the above mentioned system V794~Aql, none shows 
indications for the AH~Pic syndrome.

\subsection{Relationship to anomalous Z Cam stars}
\label{Relationship to anomalous Z Cam stars}

Another group of stars with some resemblance in their long-term light 
curves to the AH~Pic stars are the anomalous Z~Cam-type systems. As is
well known, Z~Cam stars are dwarf novae that after an outburst 
occasionally remain at an 
intermediate brightness level for an unpredictable time before returning
to the quiescent state. As first noticed by  \citet{Simonsen11}, in the 
anomalous Z~Cam stars -- a term introduced by \citet{Szkody13} -- the
standstills are instead terminated by another outburst. \citet{Kato19},
who dubbed these systems IW~And stars after their prototype, provided a
quite specific definition for the IW~And phenomenon, namely a sequence of
``standstill -- brightening -- damping oscillation''. However, soon afterwards
\citet{Kimura20b} introduced a much broader definition, saying ``it is 
another essential feature of the light variations in IW~And stars that 
they would alternate in one object between the IW~And-type phenomenon,
Z~Cam-type standstills, normal dwarf nova outbursts and heartbeat oscillations
on time scales of $\sim$100 -- 1000 days'', opening the door to classify
many more systems as IW~And stars than the much narrower original definition.
Considering the lack of well-expressed standstills and damping oscillations,
the AH~Pic stars do not fit the definition of \citet{Kato19}, but may fit 
into the
broader definition of \citet{Kimura20b} since the continuous sequence of small
outbursts can be equated to what they call ``heartbeat oscillations''. A 
detailed comparison between
the properties of AH~Pic stars and IW~And stars is therefore appropriate.

I identified in the contemporaneous literature 14 systems classified as 
IW~And stars. They are listed together with the AH~Pic stars in 
Table~\ref{Table: Comparison AH Pic - IW And}, where the most relevant
references are cited in the last column.
Several more (the AH~Pic star V2738~Ori among these) are mentioned in 
footnote~11 of \citet{Lee21}. But since these are not well documented I
will ignore them here. 

Because in two systems classified as IW~And stars -- IM~Eri 
\citep{Armstrong13} and KIC~9406652 \citep{Gies13} -- negative superhumps (nSHs)
had been observed, \citet{Kimura20b} tried with mixed success to explain
their properties within a tilted disk scenario. An assessment of the
presence of nSHs in all IW~And and AH~Pic stars based on information scattered
in the literature and in TESS and Kepler data downloaded from MAST showed
that in about half of the members of both types nSHs have been seen at some
time or the other, but in no case are these permanent\footnote{No 
observations suitable for the detection of superhumps are available for 
PS~Eri and BO~Cet. The long TESS light
curves of Karachurin~12 almost always contain nSHs, except for short time
intervals.} (second column of Table~\ref{Table: Comparison AH Pic - IW And}).
Moreover, there is no correlation between time intervals with active nSHs and 
the occurrence of the IW~And phenomenon or the AH~Pic syndrome. The power 
spectra of the TESS light curves of some of the anomalous Z~Cam stars contain 
noteworthy features not directly related to the topic of this study. These
are briefly presented in 
Appendix~\ref{Frequency analysis of some anomalous Z Cam stars}.

%-------------------------------------------------------------
\begin{table*}
\caption{Comparison between the properties of AH~Pic stars and IW~And stars}
\label{Table: Comparison AH Pic - IW And}
\centering

\begin{tabular}{lcccc}
\hline \hline
Name & nSH$^a$ & Duty     & Dip           & Ref$^d$\\
     &         & cycle$^b$ & occurence$^c$ &        \\
\hline
AH Pic stars                 &
                             & 
                             & 
                             & 
                             \\
\phantom{1ex} V1116 Cep      & % Name
+                            & % negative superhump 
0.90                         & % average outburst duty cycle in active phases
0.02                         & % fraction of outburst with dips
1                            \\ % References
\phantom{1ex} CM Del         & % Name
--                           & % negative superhump 
0.96                         & % average outburst duty cycle in active phases
0.06                         & % fraction of outburst with dips
1                            \\ % References
\phantom{1ex} PS Eri         & % Name
?                            & % negative superhump 
0.88                         & % average outburst duty cycle in active phases
0.17                         & % fraction of outburst with dips
1                            \\ % References
\phantom{1ex} V2837 Ori      & % Name
+                            & % negative superhump 
0.77                         & % average outburst duty cycle in active phases
0.05                         & % fraction of outburst with dips
1                            \\ % References
\phantom{1ex} FY Per         & % Name
+                            & % negative superhump 
0.90                         & % average outburst duty cycle in active phases
0.03                         & % fraction of outburst with dips
1                            \\ % References
\phantom{1ex} AH Pic         & % Name
--                           & % negative superhump 
0.90                         & % average outburst duty cycle in active phases
0.03                         & % fraction of outburst with dips
1                            \\ % References
\phantom{1ex} KIC 9202990    & % Name
--                           & % negative superhump 
0.90                         & % average outburst duty cycle in active phases
0.01                         & % fraction of outburst with dips
1                            \\ % References
IW And stars                 &
                             & 
                             & 
                             & 
                             \\
\phantom{1ex} IW And         & % Name
--                           & % negative superhumps
0.43                         & % average outburst duty cycle in active phases
0.74                         & % fraction of outburst with dips
2                            \\ % References
\phantom{1ex} BC Cas         & % Name
--                           & % negative superhump 
0.35                         & % average outburst duty cycle in active phases
0.33                         & % fraction of outburst with dips
3                            \\ % References
\phantom{1ex} V513 Cas       & % Name
+                            & % negative superhump 
0.45                         & % average outburst duty cycle in active phases
0.23                         & % fraction of outburst with dips
2                            \\ % References
\phantom{1ex} BO Cet         & % Name
?                            & % negative superhump 
0.64                         & % average outburst duty cycle in active phases
0.42                         & % fraction of outburst with dips
4                            \\ % References
\phantom{1ex} ST Cha         & % Name
--                           & % negative superhump 
0.48                         & % average outburst duty cycle in active phases
0.78                         & % fraction of outburst with dips
5                            \\ % References
\phantom{1ex} V507 Cyg       & % Name
+                            & % negative superhump 
0.61                         & % average outburst duty cycle in active phases
0.28                         & % fraction of outburst with dips
6,7                          \\ % References
\phantom{1ex} IM Eri         & % Name
--                           & % negative superhump 
0.34                         & % average outburst duty cycle in active phases
0.90                         & % fraction of outburst with dips
7,8                          \\ % References
\phantom{1ex} V523 Lyr       & % Name
+                            & % negative superhump 
0.33                         & % average outburst duty cycle in active phases
0.38                         & % fraction of outburst with dips
9                            \\ % References
\phantom{1ex} HO Pup         & % Name
--                           & % negative superhump 
0.57                         & % average outburst duty cycle in active phases
0.33                         & % fraction of outburst with dips
10                           \\ % References
\phantom{1ex} FY Vul         & % Name
+                            & % negative superhump 
0.64                         & % average outburst duty cycle in active phases
0.28                         & % fraction of outburst with dips
6                            \\ % References
\phantom{1ex} LAMOST J065237.19+243622.1         & % Name
+                            & % negative superhump 
0.60                         & % average outburst duty cycle in active phases
0.10                         & % fraction of outburst with dips
11                           \\ % References
\phantom{1ex} ASAS J071404+7004.3   & % Name
+                            & % negative superhump 
0.87                         & % average outburst duty cycle in active phases
0.02                         & % fraction of outburst with dips
12                           \\ % References
\phantom{1ex} KIC 9406652    & % Name
+                            & % negative superhump 
0.33                         & % average outburst duty cycle in active phases
0.56                         & % fraction of outburst with dips
13,14                        \\ % References
\phantom{1ex} Karachurin 12  & % Name
+                            & % negative superhump 
0.65                         & % average outburst duty cycle in active phases
0.03                         & % fraction of outburst with dips
15                           \\ % References
\hline
\end{tabular}
\\ 
$^a$: negative superhumps occurrence: + (sometimes), -- (never) \\
$^b$: average outburst duty cycle during active intervals \\
$^c$: fraction of outbursts followed by dips \\
$^d$: References to detailed paper:
(1) this work,
(2) \citet{Szkody13},
(3) \citet{Kato20a},
(4) \citet{Kato21},
(5) \citet{Simonsen14},
(6) \citet{Kato19},
(7) \citet{Kato22a},
(8) \citet{Kato20b},
(9) \citet{Mason16},
(10) \citet{Lee21},
(11) \citet{Sun24b},
(12) \citet{Kato22b},
(13) \citet{Gies13},
(14) \citet{Kimura20a},
(15) \citet{Sun24a}
\end{table*}
%------------------------------------------------------------------
 
Phenomenologically, the 
light curves of anomalous Z~Cam stars differ from AH~Pic type light curves
in two major aspects: (1) There is almost always an appreciable quiescent
(or standstill) interval between outbursts, while stunted outbursts in 
AH~Pic stars occur in immediate succession or almost so, and (2) most of 
the time they exhibit a dip after the outburst, while this is rather the 
exception in AH~Pic stars. 

In order to quantify these issues, the length of the standstill or quiescent 
intervals and the outburst durations were measured in the long-term 
ASAS-SN light curves (AAVSO, TESS and Kepler light curves for V1116~Cep, 
AH~Pic and KIC~9202990, respectively) of all IW~And and AH~Pic 
stars\footnote{Since no suitable long-term light curve of BC~Cas is
available, the short light curve reproduced by \citet{Kato20b} was used for
a rough estimate.}. Due 
to the sometimes insufficient sampling of these data sets it is not always 
easy to define the start and end points of these intervals or of the 
outbursts. Moreover, frequently it is not evident whether
an more modest initial increase in brightness, before a steeper rise sets in,
should be regarded as part of the outburst or as part of the preceding 
standstill. Another concern are successive local minima after an outburst,
more often seen in IW~And stars than in AH~Pic systems (``damping 
oscillations''). Should these be considered as part of the inter-outburst
(standstill) intervals or not? Such problems render an unbiased determination
of the duration of quiescent and outburst phases difficult. However, I tried
to apply the same definition to all light curves so that systematic errors
should be similar in all systems and the results should at least
roughly be comparable.

Similarly, the definition of dips after outbursts is somewhat subjective. 
Here, I consider as dips only minima that have a depth beneath the quiescent
or standstill magnitude level similar or larger than the average outburst
amplitude, that is, minima similar to those occurring during the last but one
observing season of CM~Del (upper panel of Fig.~\ref{cmdel-lc}) or occasionally
seen in PS~Eri (Fig.~\ref{j0229-3958-lc}). Not unfrequently,
in some IW~And stars such deep dips occur after successive brightenings without
preceding standstills. During such intervals the outburst activity is 
indistinguishable from normal dwarf nova behavior with rapidly repeating
outbursts. Here, they are not considered to be part of the IW~And activity. 

The average ratio between the outburst duration and the duration of an entire 
quiescence (standstill) -- outburst cycle is the outburst duty cycle listed in 
column~3 of Table~\ref{Table: Comparison AH Pic - IW And}. Column~4 contains 
the fraction of brightenings that are followed by a dip in the light curve.
Concerning both criteria there is a rather clear distinction between AH~Pic
stars and IW~And stars. In the former group the outburst duty cycle is much
higher and varies over a smaller range that in the latter. The contrast
between the groups concerning the fraction of dip occurrences is just as
striking. There is only one outlier among the IW~And stars. 
ASAS~J071404+70043 has a high duty cycle, and the dip
occurrence rate is low. This is the normal behavior of AH~Pic stars. 
Therefore, this system should better be considered as a AH~Pic star instead
of a IW~And star.

There are thus quantitative differences between both groups of stars.
Additionally, the AH~Pic stars form a significantly more homogeneous group 
compared to the anomalous Z~Cam stars, spanning a much smaller parameter
range. While some ideas about the mechanism
causing the IW~And phenomenon float in the literature 
\citep{Hameury14, Kimura20a, Kimura20b}
this question cannot be considered as solved. The reasons for the AH~Pic 
syndrome are equally unknown (see Sect.~\ref{Discussion}). Unless
the qualitatively similar, although quantitatively distinct 
phenomenological behavior of the two groups can eventually be traced
to different physical processes the possibility of a close relationship
between them remains open. The AH~Pic syndrome may then be seen as an
extreme manifestation of the IW~And phenomenon.

\section{Discussion}
\label{Discussion}

Small scale stunted outbursts in novalike variables have long been known.
Even the occurrence of long trains of such brightenings has
been observed in at least one system (FY~Per) already long ago
\citep{Honeycutt01, Honeycutt04}. A second object with similar 
properties (KIC~9202990) was identified by \citet{Ostensen10} and
\citet{Ramsay16}. Recently, yet another system (V1116~Cep)
has been recognized to show a very similar behavior 
\citep{Shears20, Shears23}. But no connexion between the three stars was
established. However, with four more NLs (CM~Del, AH~Pic, V2837~Ori and
PS~Eri) 
exhibiting the same phenomenon, distinct from almost all other systems known 
to suffer stunted outbursts, and with other strikingly similar 
characteristics, in particular the almost identical orbital waveforms, it is
appropriate to look for a common origin for this behavior. 

The question is thus: what causes the quasi-periodic
brightenings and why are they only observed in a minority of NLs?
The answer will only be possible after the mechanism leading to stunted
outbursts in the first place is identified. Only then more specific
details such as the cause for the slow evolution of outburst intervals
and profiles, and their correlation, or the reason for the temporary
suspension of the outburst activity can be addressed. 
It appears that we are still
far from understanding this phenomenon. Already early-on \citet{Honeycutt98a}
discussed arguments in favour of possible mechanisms, as well as their 
problems. These mechanisms can basically be divided into two categories,
namely (1) modulation of the mass transfer rate from the secondary star, and 
(2) disk instabilities similar to those leading to dwarf nova outbursts, also
considering truncated accretion disks and Z~Cam-like outbursts. However,
for none of the scenarios a more specific mechanism has convincingly been 
proposed. The lack of a more detailed understanding of the underlying 
physics is discussed by \citet{Hameury20}. Subsequently, I will regard
in some more detail the arguments in favour and the problems of the two
scenarios with particular emphasize on the AH~Pic syndrome.

\subsection{Disk instabilities}
\label{Disk instabilities}

Ordinary dwarf nova outbursts have extensively been investigated 
within the DIM theory \citep[see][for a thorough review]{Lasota01}. 
Even if many questions concerning specific aspects are still open, 
quasi-periodic outbursts, including events occurring in quick succession, have 
routinely been simulated. Were it not for the small amplitude, the outburst
behavior of the AH~Pic stars would not be unlike that of dwarf novae such as 
CN~Ori which has a very high duty cycle as evidenced by its ASAS-SN light 
curve. 

One of the problems often mentioned in this context concerns the average 
mass transfer rate $\dot{M}$ from the secondary. In order for the DIM to 
work it must be below a given threshold $\dot{M}_{\rm crit}$. However, 
novalike variables are generally
supposed to have a high $\dot{M}$ which keeps the accretion disk in a
permanent high state not unlike dwarf novae during outbursts. Could it
be that systems with stunted outbursts and in particular the AH~Pic
stars have mass transfer rates close to $\dot{M}_{\rm crit}$ so that
disk instabilities are still enabled? For none of the AH~Pic stars $\dot{M}$
has been measured. But their absolute magnitude may serve as a rough 
proxy. As shown in Fig.~\ref{Mv-distr}
they are all in the range where the distribution function of NLs
and dwarf novae overlap. This may be an indication that the mass
transfer rate through their disks is within the range that still permits
disk instabilities and dwarf novae outbursts. 

But then, the likelihood for stunted outbursts to occur should be enhanced
when the overall brightness of the system is reduced, that is, when the
average mass transfer rate is smaller. However, 
\citet{Honeycutt01} concludes that there is no correlation between the 
occurrence of stunted outbursts and the system brightness either in a given 
system or from system to system. Moreover, two distinct minima can be seen 
in the long-term light curve of V1116~Cep (Fig.~\ref{hs0229-lc}), one close 
to JD 2456000, the other around JD 2457750. During both events outbursts are 
suppressed. Similarly, during the low state of the candidate system V794~Aql 
the outburst activity is at least temporarily suspended. 
There may thus be a tendency for stunted outbursts to occur when the overall 
system brightness is higher. This goes in the wrong direction if stunted 
outbursts were to be explained within the DIM scenario.

\citet{Honeycutt98a} raised the hypothesis that stunted outbursts are just
like normal dwarf novae outbursts but seen against an attenuating background 
light source. \citet{Honeycutt01} discussed this idea further.
He compared the amplitude distribution of stunted outbursts with
that of SS~Cyg- and Z~Cam-type dwarf novae derived from compilations of 
\citet{Warner95a} and found the best congruity if the diluting light source
is 3.5~mag (or 25 times)  brighter than the outbursting light source in
quiescence. The data listed in 
Table~\ref{Table: Dwarf novae absolute magnitudes} 
yield a mean outburst amplitude of $2.5 \pm 1.0$~mag, somewhat less than
the center of the distribution used by \citet{Honeycutt01}. Together with
the mean amplitude of the outbursts of the AH~Pic stars of
$0.74 \pm 0.18$~mag the constant light should then on average be 
% $8.3744 \pm 9.9721$
8.4 times (2.3~mag) brighter than the variable light at minimum. Thus, the
contrast between the two light sources needs not to be as drastic as the
3.5~mag found by \citet{Honeycutt01}. But still, although \citet{Honeycutt01}
briefly mentions some possible scenarios it is not obvious where the diluting 
light should come from. 

Another hypothesis discussed by \citet{Honeycutt98a} concerns the idea that
small amplitude outburst caused by the DIM mechanism may be possible if the
accretion disk is truncated and the inner parts of the disk are missing.
A disk may be truncated for different reasons. The most widely investigated
possibility is a disruption of the inner disk by the strong magnetic field
in intermediate polars. In that case a light curve modulation on time scales
of minutes is most often visible, revealing the rotation period of the
white dwarf. This effect is easily detected in TESS light curves (Bruch, 
in preparation), but it is not seen in the TESS data of any of the AH~Pic
stars. Therefore, a disk truncated by a white dwarf magnetic field cannot
explain the AH~Pic syndrome, even more so considering that outbursts in 
truncated disks are supposed to be infrequent, as \citet{Honeycutt98a} 
pointed out.

Dwarf nova outbursts occur semi-periodically. While this is also the case
for the AH~Pic stars, it is not so for the majority of stars exhibiting
stunted outbursts where the brightenings occur quite irregularly or only
sporadically, sometimes in small clusters separated by long intervals 
without outburst activity. Thus, if stunted outbursts are indeed caused
by disk instabilities, an explanation for the quite distinct outburst
pattern in AH-Pic stars and in other novalike variables is required.

\subsection{Mass transfer variations}
\label{Mass transfer variations}

An alternative explanation for stunted outbursts are mass transfer bursts
from the secondary star. Some observations appear to favor this mechanism, 
but there are also problems. 

\citet{Baptista11}, using eclipse mapping techniques, find that the behavior
of UU~Aqr during a stunted outburst is incompatible with the disk instability
model for dwarf nova outbursts and therefore prefer modulations of the mass
transfer to explain the brightenings. While this is not a direct argument in
favor of mass transfer variations, \citet{Robertson18} defend this notion
even more emphatically. The principle diagnostic tool is the hot spot,
that is, the region where the stream of transferred matter hits the
accrection disk edge. Variations of the mass transfer rate should lead
to a modulation of the brightness of the hot spot, and -- in eclipsing
systems -- to variations of the eclipse shape and the minimum phase. 
\citet{Robertson18} analysed eclipse profiles in UU~Aqr during stunted 
outbursts and found in about half of them hot spot enhancements and a delayed 
eclipse egress. They take this as evidence for increased mass transfer. They 
hypothesize that the absence of such signatures in the remaining eclipse 
profiles may be due to blobby mass transfer, such that the stream of matter 
is not constant on the orbital time scale. 

A change of the eclipse profiles as a function of the outburst phase is also
evident in the two eclipsing AH~Pic stars CM~Del and KIC~9202990
(Fig.~\ref{waveforms}). In a 
symmetrical accretion disk the light center will coincide with the disk 
center and thus the white dwarf location. A hot spot will shift the light 
center away from the disk center, and this effect will increase with the
hot spot brightness. Moreover, the light center will become broader.
Consequently, the eclipse profile will also broaden, and the minimum
will be shifted in phase (to later phases because the hot spot is eclipsed
after the disk center). In CM~Del and KIC~9202990, a shift of 0.030 and 0.029,
respectively, of the minimum phase is observed between the faint and bright
outburst phases. Moreover, at least in KIC~9202990 the eclipse also appears to 
be broader during outburst maximum. 

On the other hand, from Fig.~2 of \citet{Smak89} it can be estimated that
a mass transfer enhancement of the order of 10 times is required for an
increase of the disk brightness by 0.5 -- 1~mag, typical for stunted outbursts.
As \citet{Warner95b} pointed out, in dwarf novae (having fainter accretion
disks than novalike variables but similar secondary stars) mass transfer 
modulations of this magnitude would cause strong variations of the hot spot 
brightness which are not observed on time scales of stunted outbursts. This
constitutes a problem for scenarios to explain stunted outbursts by mass
transfer variations from the secondary stars. Moreover, concerning the AH~Pic 
syndrome, a mechanism would be required to drive such variations over years
semi-periodically on time scales of week. I am not aware of
such a mechanism. The rather exotic idea that a third body in an eccentric
orbit around the central binary system may induce slight variations of
the Roche lobe size of the secondary star and thus variations of its
atmospheric scale height at the $L_1$ point, causing the mass transfer to
increase and decrease, can be refuted because this effect would be strictly
periodic, while the outburst intervals in AH~Pic stars are not stable.

\subsection{Dips}
\label{Dips}

\begin{figure}
\figurenum{16}
\label{j0229-3958-lc-detail}
\plotone{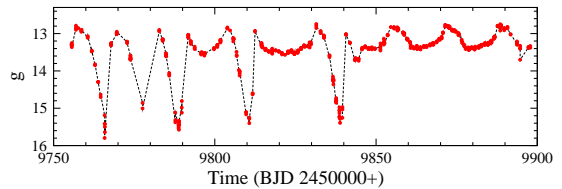}
\caption{Detail of the ASAS-SN light curve of PS~Eri containing
         dips and the transition to the normal outburst behavior.} 
\end{figure}

An enigmatic feature associated with stunted outbursts are brightness dips 
which occur occasionally right after the end of the outburst. This happens
in systems with sporadic outbursts \citep{Honeycutt98a, Honeycutt01} as
well as in some AH~Pic stars (CM~Del: Fig.~\ref{cmdel-lc};
AH~Pic: Fig.~\ref{ahpic-lc}; KIC~9202990: Fig.~\ref{kic9202990-lc}) and
is particularly evident in PS~Eri (Fig.~\ref{j0229-3958-lc}). 
There is a strong tendency for these events to cluster within limited
time intervals. Such dips occur even more frequently in the anomalous
Z~Cam stars (see Sect.~\ref{Relationship to anomalous Z Cam stars}), but
dip clustering is much less evident in these systems.
An amplified section of the ASAS-SN light curve of 
PS~Eri containing some dips and the transition to the 
normal outburst behavior is shown Fig.~\ref{j0229-3958-lc-detail}.

During the dip intervals the total amplitude of the variations becomes
similar to the average outburst amplitudes of dwarf novae. As a curious
feature I note that, while the limited
data sampling precludes a corresponding statement for the other dipping
AH~Pic stars, the dips (at least the last three dips in 
Fig.~\ref{j0229-3958-lc-detail}) in PS~Eri appear not to
occur after a stunted outburst, but interrupt the outburst. That is, the
outburst is still in progress after recovery from the dip. I am not aware
that this strange behavior has been observed before in a stunted outburst 
of any NL. But a similar effect can be identified in the ASAS-SN 
light curves of some IW~And stars.

\subsection{Some basic statistics}
\label{Some stattistics}

AH~Pic stars appear not to be numerous.
Based on his monitoring of NLs \citet{Honeycutt01} estimated that the fraction
of novalike variables exhibiting stunted outbursts could approach 50\%.
His list of positive detections contains nine systems (not counting X~Ser;
see Sect.~\ref{Related systems}). Of 13 old novae observed by \citet{Vogt18} 
six underwent such brightenings, leading to a similar fraction. Adding to
these studies the list of \citet{Zsidi23}, 24~NLs exhibit stunted outbursts.
None of these studies did specifically aim targets with continuous sequences 
of outbursts and thus should not be grossly biased towards such stars. 
Among the 24 systems only two are AH~Pic stars (CM~Del and FY~Per). Taking 
these numbers at face value, a fraction of about 
$0.5 \times 2/24 \approx 0.04$ of all novalike
variables are expected to be AH~Pic stars. Acknowledging that I am dealing
with small number statistics, this result should not be taken too seriously.
However, it shows that the AH~Pic stars are probably a small minority among 
NLs. 

\section{Summary}
\label{Summary}

New observational facilities in space and on Earth together with amateur 
astronomers having access to ever more sophisticated instrumentation
has enabled close monitoring of many variable stars during the past one
or two decades to a degree that was unthinkable previously. Light
curves with no or few interruptions extending over years, sometimes 
with a time resolution as short as a minute, enabled the detection of
never before seen features. Among these are the phenomena studied in
this paper. 

Low scale brightenings in novalike variables lasting for a week or so
have been observed before. But they were mostly seen as isolated events.
The detection of long trains of such stunted outburst in close succession 
lasting for years required the new approach mentioned above. Of the seven stars 
identified to exhibit this behavior, four (V1116~Cep, CM~Del, FY~Per, and
KIC~9202990) were noted in the past to show this property. However, it was 
never discussed as a special feature and no connexion between the stars was
established. The three additional systems presented here (AH~Pic, 
V2837~Ori, and PS~Eri), 
plus the detection of other common properties beyond the continuous stunted 
outbursts, namely the very similar stable orbital waveforms and their 
consistent variations during and between brightenings, and their absolute
magnitudes in the overlap range between absolute magnitudes of NLs and dwarf
novae above the CV period gap, justifies to 
regard them as a group. An understanding of this AH~Pic syndrome 
has not yet been achieved. It requires in the first place the identification 
of the mechanism leading to stunted outbursts, and then an explanation for 
their continuous occurrence in AH~Pic star, while they are seen only 
occasionally or not at all in the large majority of novalike variables. 
Only then more specific details of their behavior can be addressed,

\begin{acknowledgements}
This paper is based on data collected by the TESS and Kepler missions and 
obtained from the MAST data archive at the Space Telescope Science Institute 
(STScI). Funding for the missions is provided by the NASA Explorer Program
and the NASA Science Mission Directorate for TESS and Kepler, respectively. 
STScI is operated by the Association of Universities for Research in 
Astronomy, Inc., under NASA contract NAS 5-26555. Further data collected
by ASAS-SN are used. This project is funded in part by the Gordon and Betty 
Moore Foundation through grants GBFM5490 and GBFM10501 to the Ohio State 
University, and also in part by the Alfred P.\ Sloan Foundation grant 
G-2021-14192. Last not least, data downloaded from the AAVSO International
Database contributed to this paper. I am grateful to all the individuals who
keep these projects alive and to the many observers who supply data to the
AAVSO archives. I also thank the anonymous referee for critical comments 
which led to an improvement of the original version of this paper.
\end{acknowledgements}

\appendix

\section{Absolute magnitude distribution functions}
\label{Absolute magnitude distribution functions}

The distribution function of the absolute $V$ magnitudes of SS~Cyg and Z~Cam
type dwarf novae during quiescence and outburst maximum (i.e., avoiding the 
systematically 
fainter SU~UMa and WZ~Sge type stars) was calculated based on a sample of 
89 systems with orbital periods above the CV period gap and below 10~h. The 
average apparent magnitudes were estimated in their ASAS-SN
$V$ band light curves. Together with their Gaia DR3 distances 
\citep{Bailer-Jones21} and the interstellar absorption, calculated using the
$E_{B-V}$ values taken from the compilation of \citet{Bruch94} or estimated
from the tridimensional extinction maps 
\citep[Stilism,][]{Lallement14, Capitanio17} 
this yields the absolute magnitude. Since the
magnitude depends on the orbital inclination a correction for this effect 
using the formula of \citet{Webbink87} was applied to all systems with 
known inclination. The latter were taken from the final edition (Dezember 2016) 
of the Ritter \& Kolb catalogue \citep{Ritter03} or references cited 
therein. In a few cases of eclipsing systems without measured
inclination a rough estimate based on the eclipse depth was made. 
The results are listed in Table~\ref{Table: Dwarf novae absolute magnitudes}.

\startlongtable
\begin{deluxetable*}{lccccccc}
%\startlongtable
\tablenum{A1}
\tablecaption{Average apparent $V$ magnitude during quiescence and outburst, 
distances, interstellar
reddenings, orbital inclinations and average absolute $V$ magnitudes for 89
SS~Cyg- and Z~Cam-type dwarf novae.}
\label{Table: Dwarf novae absolute magnitudes}
%\tablewidth{0pt}
\tablehead{
\colhead{Name} &
\colhead{$m_V$} &
\colhead{$m_V$} &
\colhead{Distance} &
\colhead{$E_{B-V}$} &
\colhead{Incl.} &
\colhead{$M_V$} &
\colhead{$M_V$} \\
\colhead{} &
\colhead{(qui)} &
\colhead{(outb)} &
\colhead{pc} &
\colhead{} &
\colhead{($^o$)} &
\colhead{(qui)} &
\colhead{(outb)} \\
}
%\colnumbers
\startdata
RX And\phantom{054558.3+022106}   &  % name of object
13.4                    &  % apparent quiescent magnitude
10.9                    &  % apparent quiescent magnitude
\phantom{1}197          &  % distance (pc)
0.02                    &  % interstellar reddening E(B-V)
51                      &  % orbital inclination
7.1                     &  % absolute quiescent magnitude
4.7                     \\ % absolute maximum magnitude
AR And                  &  % name of object
16.2                    &  % apparent quiescent magnitude
12.7                    &  % apparent quiescent magnitude
\phantom{1}410          &  % distance (pc)
0.02                    &  % interstellar reddening E(B-V)
\nodata                 &  % orbital inclination
8.2                     &  % absolute quiescent magnitude
4.6                     \\ % absolute maximum magnitude
DX And                  &  % name of object
15.1                    &  % apparent quiescent magnitude
12.3                    &  % apparent quiescent magnitude
\phantom{1}585          &  % distance (pc)
0.11                    &  % interstellar reddening E(B-V)
45                      &  % orbital inclination
6.3                     &  % absolute quiescent magnitude
3.5                     \\ % absolute maximum magnitude
IW And                  &  % name of object
16.7                    &  % apparent quiescent magnitude
14.0                    &  % apparent quiescent magnitude
\phantom{1}850          &  % distance (pc)
0.06                    &  % interstellar reddening E(B-V)
\nodata                 &  % orbital inclination
6.9                     &  % absolute quiescent magnitude
4.3                     \\ % absolute maximum magnitude
LX And                  &  % name of object
15.4                    &  % apparent quiescent magnitude
13.1                    &  % apparent quiescent magnitude
\phantom{1}508          &  % distance (pc)
0.05                    &  % interstellar reddening E(B-V)
\nodata                 &  % orbital inclination
6.8                     &  % absolute quiescent magnitude
4.5                     \\ % absolute maximum magnitude
V823 And                &  % name of object
14.9                    &  % apparent quiescent magnitude
14.0                    &  % apparent quiescent magnitude
\phantom{1}469          &  % distance (pc)
0.08                    &  % interstellar reddening E(B-V)
80                      &  % orbital inclination
5.1                     &  % absolute quiescent magnitude
4.2                     \\ % absolute maximum magnitude
VZ Aqr                  &  % name of object
16.6                    &  % apparent quiescent magnitude
12.9                    &  % apparent quiescent magnitude
\phantom{1}598          &  % distance (pc)
0.03                    &  % interstellar reddening E(B-V)
\nodata                 &  % orbital inclination
7.7                     &  % absolute quiescent magnitude
4.0                     \\ % absolute maximum magnitude
UU Aql                  &  % name of object
16.3                    &  % apparent quiescent magnitude
11.7                    &  % apparent quiescent magnitude
\phantom{1}306          &  % distance (pc)
0.13                    &  % interstellar reddening E(B-V)
41                      &  % orbital inclination
9.0                     &  % absolute quiescent magnitude
4.3                     \\ % absolute maximum magnitude
FO Aql                  &  % name of object
16.1                    &  % apparent quiescent magnitude
13.9                    &  % apparent quiescent magnitude
\phantom{1}514          &  % distance (pc)
0.06                    &  % interstellar reddening E(B-V)
\nodata                 &  % orbital inclination
7.5                     &  % absolute quiescent magnitude
5.3                     \\ % absolute maximum magnitude
PQ Aql                  &  % name of object
16.3                    &  % apparent quiescent magnitude
14.7                    &  % apparent quiescent magnitude
1388                    &  % distance (pc)
0.07                    &  % interstellar reddening E(B-V)
\nodata                 &  % orbital inclination
5.4                     &  % absolute quiescent magnitude
3.9                     \\ % absolute maximum magnitude
V1101 Aql               &  % name of object
15.2                    &  % apparent quiescent magnitude
13.9                    &  % apparent quiescent magnitude
\phantom{1}687          &  % distance (pc)
0.05                    &  % interstellar reddening E(B-V)
\nodata                 &  % orbital inclination
5.9                     &  % absolute quiescent magnitude
4.7                     \\ % absolute maximum magnitude
AT Ara                  &  % name of object
14.3                    &  % apparent quiescent magnitude
12.4                    &  % apparent quiescent magnitude
\phantom{1}783          &  % distance (pc)
0.12                    &  % interstellar reddening E(B-V)
38                      &  % orbital inclination
5.0                     &  % absolute quiescent magnitude
3.0                     \\ % absolute maximum magnitude
SS Aur                  &  % name of object
14.6                    &  % apparent quiescent magnitude
11.2                    &  % apparent quiescent magnitude
\phantom{1}249          &  % distance (pc)
0.05                    &  % interstellar reddening E(B-V)
38                      &  % orbital inclination
7.9                     &  % absolute quiescent magnitude
4.6                     \\ % absolute maximum magnitude
Z Cam                   &  % name of object
13.1                    &  % apparent quiescent magnitude
10.5                    &  % apparent quiescent magnitude
\phantom{1}213          &  % distance (pc)
0.02                    &  % interstellar reddening E(B-V)
57                      &  % orbital inclination
6.5                     &  % absolute quiescent magnitude
3.9                     \\ % absolute maximum magnitude
AF Cam                  &  % name of object
16.9                    &  % apparent quiescent magnitude
13.6                    &  % apparent quiescent magnitude
\phantom{1}874          &  % distance (pc)
0.32                    &  % interstellar reddening E(B-V)
\nodata                 &  % orbital inclination
6.3                     &  % absolute quiescent magnitude
3.0                     \\ % absolute maximum magnitude
SY Cnc                  &  % name of object
13.3                    &  % apparent quiescent magnitude
11.2                    &  % apparent quiescent magnitude
\phantom{1}401          &  % distance (pc)
0.00                    &  % interstellar reddening E(B-V)
60                      &  % orbital inclination
5.3                     &  % absolute quiescent magnitude
3.2                     \\ % absolute maximum magnitude
AT Cnc                  &  % name of object
14.7                    &  % apparent quiescent magnitude
12.7                    &  % apparent quiescent magnitude
\phantom{1}455          &  % distance (pc)
0.03                    &  % interstellar reddening E(B-V)
\nodata                 &  % orbital inclination
6.4                     &  % absolute quiescent magnitude
4.5                     \\ % absolute maximum magnitude
GY Cnc                  &  % name of object
16.3                    &  % apparent quiescent magnitude
12.9                    &  % apparent quiescent magnitude
\phantom{1}271          &  % distance (pc)
0.02                    &  % interstellar reddening E(B-V)
77                      &  % orbital inclination
8.1                     &  % absolute quiescent magnitude
4.8                     \\ % absolute maximum magnitude
AM Cas                  &  % name of object
14.6                    &  % apparent quiescent magnitude
13.0                    &  % apparent quiescent magnitude
\phantom{1}421          &  % distance (pc)
0.12                    &  % interstellar reddening E(B-V)
\nodata                 &  % orbital inclination
6.2                     &  % absolute quiescent magnitude
4.6                     \\ % absolute maximum magnitude
HL CMa                  &  % name of object
13.2                    &  % apparent quiescent magnitude
11.1                    &  % apparent quiescent magnitude
\phantom{1}294          &  % distance (pc)
0.06                    &  % interstellar reddening E(B-V)
45                      &  % orbital inclination
6.0                     &  % absolute quiescent magnitude
3.9                     \\ % absolute maximum magnitude
SV CMi                  &  % name of object
15.9                    &  % apparent quiescent magnitude
13.1                    &  % apparent quiescent magnitude
\phantom{1}414          &  % distance (pc)
0.02                    &  % interstellar reddening E(B-V)
\nodata                 &  % orbital inclination
7.8                     &  % absolute quiescent magnitude
5.1                     \\ % absolute maximum magnitude
BV Cen                  &  % name of object
13.0                    &  % apparent quiescent magnitude
11.6                    &  % apparent quiescent magnitude
\phantom{1}359          &  % distance (pc)
0.10                    &  % interstellar reddening E(B-V)
53                      &  % orbital inclination
5.1                     &  % absolute quiescent magnitude
3.7                     \\ % absolute maximum magnitude
MU Cen                  &  % name of object
14.8                    &  % apparent quiescent magnitude
12.7                    &  % apparent quiescent magnitude
\phantom{1}505          &  % distance (pc)
0.06                    &  % interstellar reddening E(B-V)
58                      &  % orbital inclination
6.2                     &  % absolute quiescent magnitude
4.0                     \\ % absolute maximum magnitude
V442 Cen                &  % name of object
15.6                    &  % apparent quiescent magnitude
12.0                    &  % apparent quiescent magnitude
\phantom{1}343          &  % distance (pc)
0.15                    &  % interstellar reddening E(B-V)
\nodata                 &  % orbital inclination
7.6                     &  % absolute quiescent magnitude
4.0                     \\ % absolute maximum magnitude
WW Cet                  &  % name of object
13.5                    &  % apparent quiescent magnitude
11.2                    &  % apparent quiescent magnitude
\phantom{1}218          &  % distance (pc)
0.03                    &  % interstellar reddening E(B-V)
48                      &  % orbital inclination
7.0                     &  % absolute quiescent magnitude
4.7                     \\ % absolute maximum magnitude
ST Cha                  &  % name of object
15.5                    &  % apparent quiescent magnitude
13.6                    &  % apparent quiescent magnitude
\phantom{1}690          &  % distance (pc)
0.21                    &  % interstellar reddening E(B-V)
\nodata                 &  % orbital inclination
5.8                     &  % absolute quiescent magnitude
3.9                     \\ % absolute maximum magnitude
TT Crt                  &  % name of object
15.8                    &  % apparent quiescent magnitude
13.3                    &  % apparent quiescent magnitude
\phantom{1}532          &  % distance (pc)
0.03                    &  % interstellar reddening E(B-V)
\nodata                 &  % orbital inclination
7.2                     &  % absolute quiescent magnitude
4.6                     \\ % absolute maximum magnitude
SS Cyg                  &  % name of object
11.9                    &  % apparent quiescent magnitude
\phantom{1}8.9          &  % apparent quiescent magnitude
\phantom{1}113          &  % distance (pc)
0.07                    &  % interstellar reddening E(B-V)
51                      &  % orbital inclination
6.6                     &  % absolute quiescent magnitude
3.7                     \\ % absolute maximum magnitude
EM Cyg                  &  % name of object
13.3                    &  % apparent quiescent magnitude
12.3                    &  % apparent quiescent magnitude
\phantom{1}356          &  % distance (pc)
0.03                    &  % interstellar reddening E(B-V)
67                      &  % orbital inclination
5.2                     &  % absolute quiescent magnitude
4.2                     \\ % absolute maximum magnitude
V516 Cyg                &  % name of object
16.3                    &  % apparent quiescent magnitude
13.9                    &  % apparent quiescent magnitude
\phantom{1}722          &  % distance (pc)
0.48                    &  % interstellar reddening E(B-V)
\nodata                 &  % orbital inclination
5.6                     &  % absolute quiescent magnitude
3.2                     \\ % absolute maximum magnitude
V542 Cyg                &  % name of object
17.7                    &  % apparent quiescent magnitude
13.9                    &  % apparent quiescent magnitude
1482                    &  % distance (pc)
0.08                    &  % interstellar reddening E(B-V)
\nodata                 &  % orbital inclination
6.7                     &  % absolute quiescent magnitude
2.9                     \\ % absolute maximum magnitude
V792 Cyg                &  % name of object
15.9                    &  % apparent quiescent magnitude
14.2                    &  % apparent quiescent magnitude
1302                    &  % distance (pc)
0.10                    &  % interstellar reddening E(B-V)
\nodata                 &  % orbital inclination
5.1                     &  % absolute quiescent magnitude
3.4                     \\ % absolute maximum magnitude
V795 Cyg                &  % name of object
16.3                    &  % apparent quiescent magnitude
13.5                    &  % apparent quiescent magnitude
\phantom{1}656          &  % distance (pc)
0.06                    &  % interstellar reddening E(B-V)
\nodata                 &  % orbital inclination
7.2                     &  % absolute quiescent magnitude
4.3                     \\ % absolute maximum magnitude
V811 Cyg                &  % name of object
14.7                    &  % apparent quiescent magnitude
13.6                    &  % apparent quiescent magnitude
\phantom{1}513          &  % distance (pc)
0.05                    &  % interstellar reddening E(B-V)
70                      &  % orbital inclination
5.6                     &  % absolute quiescent magnitude
4.5                     \\ % absolute maximum magnitude
EZ Del                  &  % name of object
16.5                    &  % apparent quiescent magnitude
15.0                    &  % apparent quiescent magnitude
1474                    &  % distance (pc)
0.07                    &  % interstellar reddening E(B-V)
\nodata                 &  % orbital inclination
5.5                     &  % absolute quiescent magnitude
4.0                     \\ % absolute maximum magnitude
AB Dra                  &  % name of object
14.7                    &  % apparent quiescent magnitude
12.4                    &  % apparent quiescent magnitude
\phantom{1}397          &  % distance (pc)
0.10                    &  % interstellar reddening E(B-V)
\nodata                 &  % orbital inclination
6.5                     &  % absolute quiescent magnitude
4.2                     \\ % absolute maximum magnitude
CG Dra                  &  % name of object
16.8                    &  % apparent quiescent magnitude
15.6                    &  % apparent quiescent magnitude
1332                    &  % distance (pc)
0.04                    &  % interstellar reddening E(B-V)
\nodata                 &  % orbital inclination
6.2                     &  % absolute quiescent magnitude
4.9                     \\ % absolute maximum magnitude
DO Dra                  &  % name of object
15.8                    &  % apparent quiescent magnitude
12.6                    &  % apparent quiescent magnitude
\phantom{1}195          &  % distance (pc)
0.01                    &  % interstellar reddening E(B-V)
45                      &  % orbital inclination
9.7                     &  % absolute quiescent magnitude
6.4                     \\ % absolute maximum magnitude
ES Dra                  &  % name of object
16.1                    &  % apparent quiescent magnitude
14.3                    &  % apparent quiescent magnitude
\phantom{1}680          &  % distance (pc)
0.02                    &  % interstellar reddening E(B-V)
\nodata                 &  % orbital inclination
7.0                     &  % absolute quiescent magnitude
5.1                     \\ % absolute maximum magnitude
EX Dra                  &  % name of object
14.8                    &  % apparent quiescent magnitude
12.9                    &  % apparent quiescent magnitude
\phantom{1}240          &  % distance (pc)
0.02                    &  % interstellar reddening E(B-V)
84                      &  % orbital inclination
6.1                     &  % absolute quiescent magnitude
4.2                     \\ % absolute maximum magnitude
V416 Dra                &  % name of object
16.7                    &  % apparent quiescent magnitude
14.2                    &  % apparent quiescent magnitude
\phantom{1}595          &  % distance (pc)
0.06                    &  % interstellar reddening E(B-V)
\nodata                 &  % orbital inclination
7.7                     &  % absolute quiescent magnitude
5.2                     \\ % absolute maximum magnitude
AH Eri                  &  % name of object
17.6                    &  % apparent quiescent magnitude
14.4                    &  % apparent quiescent magnitude
1114                    &  % distance (pc)
0.08                    &  % interstellar reddening E(B-V)
\nodata                 &  % orbital inclination
7.2                     &  % absolute quiescent magnitude
4.0                     \\ % absolute maximum magnitude
BF Eri                  &  % name of object
14.8                    &  % apparent quiescent magnitude
12.4                    &  % apparent quiescent magnitude
\phantom{1}557          &  % distance (pc)
0.03                    &  % interstellar reddening E(B-V)
40                      &  % orbital inclination
6.4                     &  % absolute quiescent magnitude
4.0                     \\ % absolute maximum magnitude
LT Eri                  &  % name of object
17.1                    &  % apparent quiescent magnitude
15.0                    &  % apparent quiescent magnitude
\phantom{1}684          &  % distance (pc)
0.06                    &  % interstellar reddening E(B-V)
\nodata                 &  % orbital inclination
7.8                     &  % absolute quiescent magnitude
5.8                     \\ % absolute maximum magnitude
U Gem                   &  % name of object
14.2                    &  % apparent quiescent magnitude
\phantom{1}9.4          &  % apparent quiescent magnitude
\phantom{1} 93          &  % distance (pc)
0.04                    &  % interstellar reddening E(B-V)
70                      &  % orbital inclination
8.9                     &  % absolute quiescent magnitude
4.1                     \\ % absolute maximum magnitude
AH Her                  &  % name of object
13.7                    &  % apparent quiescent magnitude
11.6                    &  % apparent quiescent magnitude
\phantom{1}328          &  % distance (pc)
0.03                    &  % interstellar reddening E(B-V)
46                      &  % orbital inclination
6.4                     &  % absolute quiescent magnitude
4.3                     \\ % absolute maximum magnitude
GY Hya                  &  % name of object
15.9                    &  % apparent quiescent magnitude
14.7                    &  % apparent quiescent magnitude
\phantom{1}638          &  % distance (pc)
0.06                    &  % interstellar reddening E(B-V)
\nodata                 &  % orbital inclination
6.8                     &  % absolute quiescent magnitude
5.6                     \\ % absolute maximum magnitude
V495 Hya                &  % name of object
18.0                    &  % apparent quiescent magnitude
14.1                    &  % apparent quiescent magnitude
1467                    &  % distance (pc)
0.53                    &  % interstellar reddening E(B-V)
\nodata                 &  % orbital inclination
5.6                     &  % absolute quiescent magnitude
1.7                     \\ % absolute maximum magnitude
X Leo                   &  % name of object
15.9                    &  % apparent quiescent magnitude
12.2                    &  % apparent quiescent magnitude
\phantom{1}438          &  % distance (pc)
0.00                    &  % interstellar reddening E(B-V)
41                      &  % orbital inclination
8.2                     &  % absolute quiescent magnitude
4.4                     \\ % absolute maximum magnitude
HM Leo                  &  % name of object
17.2                    &  % apparent quiescent magnitude
13.4                    &  % apparent quiescent magnitude
\phantom{1}506          &  % distance (pc)
0.03                    &  % interstellar reddening E(B-V)
\nodata                 &  % orbital inclination
8.7                     &  % absolute quiescent magnitude
4.9                     \\ % absolute maximum magnitude
IU Leo                  &  % name of object
15.4                    &  % apparent quiescent magnitude
13.0                    &  % apparent quiescent magnitude
\phantom{1}749          &  % distance (pc)
0.03                    &  % interstellar reddening E(B-V)
\nodata                 &  % orbital inclination
6.0                     &  % absolute quiescent magnitude
3.7                     \\ % absolute maximum magnitude
RU LMi                  &  % name of object
16.6                    &  % apparent quiescent magnitude
14.4                    &  % apparent quiescent magnitude
\phantom{1}927          &  % distance (pc)
0.02                    &  % interstellar reddening E(B-V)
\nodata                 &  % orbital inclination
6.8                     &  % absolute quiescent magnitude
4.7                     \\ % absolute maximum magnitude
CY Lyr                  &  % name of object
16.0                    &  % apparent quiescent magnitude
13.4                    &  % apparent quiescent magnitude
\phantom{1}468          &  % distance (pc)
0.08                    &  % interstellar reddening E(B-V)
\nodata                 &  % orbital inclination
7.5                     &  % absolute quiescent magnitude
4.9                     \\ % absolute maximum magnitude
LL Lyr                  &  % name of object
17.3                    &  % apparent quiescent magnitude
13.7                    &  % apparent quiescent magnitude
\phantom{1}827          &  % distance (pc)
0.05                    &  % interstellar reddening E(B-V)
\nodata                 &  % orbital inclination
7.7                     &  % absolute quiescent magnitude
4.1                     \\ % absolute maximum magnitude
V363 Lyr                &  % name of object
17.7                    &  % apparent quiescent magnitude
15.8                    &  % apparent quiescent magnitude
1622                    &  % distance (pc)
0.04                    &  % interstellar reddening E(B-V)
\nodata                 &  % orbital inclination
6.6                     &  % absolute quiescent magnitude
4.7                     \\ % absolute maximum magnitude
V584 Lyr                &  % name of object
15.6                    &  % apparent quiescent magnitude
14.6                    &  % apparent quiescent magnitude
1011                    &  % distance (pc)
0.10                    &  % interstellar reddening E(B-V)
\nodata                 &  % orbital inclination
5.3                     &  % absolute quiescent magnitude
4.3                     \\ % absolute maximum magnitude
CW Mon                  &  % name of object
16.5                    &  % apparent quiescent magnitude
12.5                    &  % apparent quiescent magnitude
\phantom{1}326          &  % distance (pc)
0.02                    &  % interstellar reddening E(B-V)
\nodata                 &  % orbital inclination
9.0                     &  % absolute quiescent magnitude
5.0                     \\ % absolute maximum magnitude
V426 Oph                &  % name of object
12.7                    &  % apparent quiescent magnitude
10.9                    &  % apparent quiescent magnitude
\phantom{1}190          &  % distance (pc)
0.08                    &  % interstellar reddening E(B-V)
59                      &  % orbital inclination
6.1                     &  % absolute quiescent magnitude
4.3                     \\ % absolute maximum magnitude
BI Ori                  &  % name of object
16.6                    &  % apparent quiescent magnitude
14.2                    &  % apparent quiescent magnitude
\phantom{1}671          &  % distance (pc)
0.07                    &  % interstellar reddening E(B-V)
\nodata                 &  % orbital inclination
7.3                     &  % absolute quiescent magnitude
5.0                     \\ % absolute maximum magnitude
CN Ori                  &  % name of object
15.8                    &  % apparent quiescent magnitude
12.6                    &  % apparent quiescent magnitude
\phantom{1}344          &  % distance (pc)
0.00                    &  % interstellar reddening E(B-V)
67                      &  % orbital inclination
7.8                     &  % absolute quiescent magnitude
4.6                     \\ % absolute maximum magnitude
CZ Ori                  &  % name of object
16.2                    &  % apparent quiescent magnitude
12.5                    &  % apparent quiescent magnitude
\phantom{1}500          &  % distance (pc)
0.00                    &  % interstellar reddening E(B-V)
24                      &  % orbital inclination
8.3                     &  % absolute quiescent magnitude
4.7                     \\ % absolute maximum magnitude
BD Pav                  &  % name of object
15.0                    &  % apparent quiescent magnitude
12.8                    &  % apparent quiescent magnitude
\phantom{1}328          &  % distance (pc)
0.06                    &  % interstellar reddening E(B-V)
71                      &  % orbital inclination
6.7                     &  % absolute quiescent magnitude
4.6                     \\ % absolute maximum magnitude
GS Pav                  &  % name of object
17.0                    &  % apparent quiescent magnitude
14.8                    &  % apparent quiescent magnitude
\phantom{1}531          &  % distance (pc)
0.04                    &  % interstellar reddening E(B-V)
79                      &  % orbital inclination
7.3                     &  % absolute quiescent magnitude
5.1                     \\ % absolute maximum magnitude
RU Peg                  &  % name of object
12.1                    &  % apparent quiescent magnitude
10.3                    &  % apparent quiescent magnitude
\phantom{1}274          &  % distance (pc)
0.04                    &  % interstellar reddening E(B-V)
33                      &  % orbital inclination
5.4                     &  % absolute quiescent magnitude
3.6                     \\ % absolute maximum magnitude
HX Peg                  &  % name of object
15.3                    &  % apparent quiescent magnitude
13.1                    &  % apparent quiescent magnitude
\phantom{1}586          &  % distance (pc)
0.25                    &  % interstellar reddening E(B-V)
\nodata                 &  % orbital inclination
5.8                     &  % absolute quiescent magnitude
3.6                     \\ % absolute maximum magnitude
IP Peg                  &  % name of object
14.8                    &  % apparent quiescent magnitude
12.3                    &  % apparent quiescent magnitude
\phantom{1}140          &  % distance (pc)
0.01                    &  % interstellar reddening E(B-V)
84                      &  % orbital inclination
7.3                     &  % absolute quiescent magnitude
4.9                     \\ % absolute maximum magnitude
V513 Peg                &  % name of object
16.2                    &  % apparent quiescent magnitude
12.8                    &  % apparent quiescent magnitude
\phantom{1}396          &  % distance (pc)
0.05                    &  % interstellar reddening E(B-V)
\nodata                 &  % orbital inclination
8.1                     &  % absolute quiescent magnitude
4.8                     \\ % absolute maximum magnitude
TZ Per                  &  % name of object
14.2                    &  % apparent quiescent magnitude
12.8                    &  % apparent quiescent magnitude
\phantom{1}457          &  % distance (pc)
0.27                    &  % interstellar reddening E(B-V)
\nodata                 &  % orbital inclination
5.1                     &  % absolute quiescent magnitude
3.8                     \\ % absolute maximum magnitude
FO Per                  &  % name of object
16.2                    &  % apparent quiescent magnitude
13.5                    &  % apparent quiescent magnitude
\phantom{1}588          &  % distance (pc)
0.00                    &  % interstellar reddening E(B-V)
\nodata                 &  % orbital inclination
7.5                     &  % absolute quiescent magnitude
4.8                     \\ % absolute maximum magnitude
KT Per                  &  % name of object
14.3                    &  % apparent quiescent magnitude
12.2                    &  % apparent quiescent magnitude
\phantom{1}245          &  % distance (pc)
0.18                    &  % interstellar reddening E(B-V)
\nodata                 &  % orbital inclination
6.9                     &  % absolute quiescent magnitude
4.8                     \\ % absolute maximum magnitude
BV Pup                  &  % name of object
15.1                    &  % apparent quiescent magnitude
13.2                    &  % apparent quiescent magnitude
\phantom{1}563          &  % distance (pc)
0.04                    &  % interstellar reddening E(B-V)
\nodata                 &  % orbital inclination
6.4                     &  % absolute quiescent magnitude
4.4                     \\ % absolute maximum magnitude
BX Pup                  &  % name of object
16.2                    &  % apparent quiescent magnitude
14.4                    &  % apparent quiescent magnitude
\phantom{1}739          &  % distance (pc)
0.05                    &  % interstellar reddening E(B-V)
\nodata                 &  % orbital inclination
6.8                     &  % absolute quiescent magnitude
5.0                     \\ % absolute maximum magnitude
V729 Sgr                &  % name of object
16.0                    &  % apparent quiescent magnitude
14.1                    &  % apparent quiescent magnitude
\phantom{1}448          &  % distance (pc)
0.10                    &  % interstellar reddening E(B-V)
\nodata                 &  % orbital inclination
7.5                     &  % absolute quiescent magnitude
5.6                     \\ % absolute maximum magnitude
RY Ser                  &  % name of object
15.9                    &  % apparent quiescent magnitude
13.4                    &  % apparent quiescent magnitude
\phantom{1}629          &  % distance (pc)
0.29                    &  % interstellar reddening E(B-V)
\nodata                 &  % orbital inclination
6.1                     &  % absolute quiescent magnitude
3.6                     \\ % absolute maximum magnitude
UZ Ser                  &  % name of object
16.4                    &  % apparent quiescent magnitude
12.9                    &  % apparent quiescent magnitude
\phantom{1}306          &  % distance (pc)
0.25                    &  % interstellar reddening E(B-V)
\nodata                 &  % orbital inclination
8.3                     &  % absolute quiescent magnitude
4.8                     \\ % absolute maximum magnitude
VZ Sex                  &  % name of object
16.7                    &  % apparent quiescent magnitude
12.8                    &  % apparent quiescent magnitude
1498                    &  % distance (pc)
0.03                    &  % interstellar reddening E(B-V)
\nodata                 &  % orbital inclination
5.8                     &  % absolute quiescent magnitude
1.9                     \\ % absolute maximum magnitude
TW Tri                  &  % name of object
16.3                    &  % apparent quiescent magnitude
15.6                    &  % apparent quiescent magnitude
\phantom{1}684          &  % distance (pc)
0.06                    &  % interstellar reddening E(B-V)
\nodata                 &  % orbital inclination
7.1                     &  % absolute quiescent magnitude
6.3                     \\ % absolute maximum magnitude
CH UMa                  &  % name of object
15.1                    &  % apparent quiescent magnitude
10.8                    &  % apparent quiescent magnitude
\phantom{1}374          &  % distance (pc)
0.02                    &  % interstellar reddening E(B-V)
21                      &  % orbital inclination
7.8                     &  % absolute quiescent magnitude
3.6                     \\ % absolute maximum magnitude
TW Vir                  &  % name of object
16.9                    &  % apparent quiescent magnitude
12.2                    &  % apparent quiescent magnitude
\phantom{1}400          &  % distance (pc)
0.03                    &  % interstellar reddening E(B-V)
43                      &  % orbital inclination
9.2                     &  % absolute quiescent magnitude
4.5                     \\ % absolute maximum magnitude
HS 0218+3229            &  % name of object
16.2                    &  % apparent quiescent magnitude
12.4                    &  % apparent quiescent magnitude
\phantom{1}486          &  % distance (pc)
0.08                    &  % interstellar reddening E(B-V)
59                      &  % orbital inclination
7.6                     &  % absolute quiescent magnitude
3.7                     \\ % absolute maximum magnitude
CRTS J044902.7-184129   &  % name of object
16.9                    &  % apparent quiescent magnitude
14.4                    &  % apparent quiescent magnitude
\phantom{1}898          &  % distance (pc)
0.03                    &  % interstellar reddening E(B-V)
65                      &  % orbital inclination
6.9                     &  % absolute quiescent magnitude
4.4                     \\ % absolute maximum magnitude
CRTS J054558.3+022106   &  % name of object
17.1                    &  % apparent quiescent magnitude
14.7                    &  % apparent quiescent magnitude
\phantom{1}532          &  % distance (pc)
0.30                    &  % interstellar reddening E(B-V)
84                      &  % orbital inclination
5.8                     &  % absolute quiescent magnitude
3.4                     \\ % absolute maximum magnitude
CRTS J064729.3+495027   &  % name of object
16.4                    &  % apparent quiescent magnitude
13.3                    &  % apparent quiescent magnitude
1011                    &  % distance (pc)
0.08                    &  % interstellar reddening E(B-V)
\nodata                 &  % orbital inclination
6.3                     &  % absolute quiescent magnitude
3.1                     \\ % absolute maximum magnitude
SDSS J090016.56+430118.2&  % name of object
17.9                    &  % apparent quiescent magnitude
14.4                    &  % apparent quiescent magnitude
\phantom{1}852          &  % distance (pc)
0.03                    &  % interstellar reddening E(B-V)
\nodata                 &  % orbital inclination
8.3                     &  % absolute quiescent magnitude
4.7                     \\ % absolute maximum magnitude
CRTS J090210.2-113032   &  % name of object
16.9                    &  % apparent quiescent magnitude
13.4                    &  % apparent quiescent magnitude
\phantom{1}885          &  % distance (pc)
0.04                    &  % interstellar reddening E(B-V)
\nodata                 &  % orbital inclination
7.1                     &  % absolute quiescent magnitude
3.7                     \\ % absolute maximum magnitude
RX J1831.7+6511         &  % name of object
16.6                    &  % apparent quiescent magnitude
14.2                    &  % apparent quiescent magnitude
\phantom{1}854          &  % distance (pc)
0.04                    &  % interstellar reddening E(B-V)
\nodata                 &  % orbital inclination
6.9                     &  % absolute quiescent magnitude
4.6                     \\ % absolute maximum magnitude
CRTS J210846.4-035031   &  % name of object
16.2                    &  % apparent quiescent magnitude
15.3                    &  % apparent quiescent magnitude
\phantom{1}756          &  % distance (pc)
0.04                    &  % interstellar reddening E(B-V)
75                      &  % orbital inclination
6.0                     &  % absolute quiescent magnitude
5.0                     \\ % absolute maximum magnitude
HS 2325+8205            &  % name of object
16.6                    &  % apparent quiescent magnitude
14.0                    &  % apparent quiescent magnitude
\phantom{1}464          &  % distance (pc)
0.12                    &  % interstellar reddening E(B-V)
\nodata                 &  % orbital inclination
8.0                     &  % absolute quiescent magnitude
5.4                     \\ % absolute maximum magnitude
Lanning 386             &  % name of object
15.4                    &  % apparent quiescent magnitude
14.9                    &  % apparent quiescent magnitude
\phantom{1}572          &  % distance (pc)
0.06                    &  % interstellar reddening E(B-V)
\nodata                 &  % orbital inclination
6.5                     &  % absolute quiescent magnitude
6.0                     \\ % absolute maximum magnitude
\enddata
%\endlongtable
\end{deluxetable*}

This approach contains several error sources. Flickering, orbital
variations, insufficent data sampling, measurement errors and variations on
long time scales all contribute to difficulties in defining the quiescent 
magnitude level in the ASAS-SN light curves which therefore is often uncertain
by up to some tenths of a magnitude. Some light curves contain only few 
outbursts and due to the sampling of the observations the maximum may have
been missed. The interstellar extinction also has 
considerable uncertainties, noting that in some cases the Stilism maps do
not extend to the distance of the targets and thus provide only lower limits.
Additionally, uncertain inclinations or the absence of a correponding 
correction in the cases with unknown inclination contribute to errors of 
the absolute magnitude. Compared to these effects, distance errors 
introduce only moderate uncertainties. Note also, that no correction for 
the contribution of the secondary star was applied which may be substantial 
in the longer period systems. Overall, the individual absolute $V$ magnitudes 
listed in Table~\ref{Table: Dwarf novae absolute magnitudes}
should be taken with a grain of salt. However, since 
most of the error sources are statistical and not systematic, they should 
cancel out in the ensemble average and thus not change substantially the 
distribution function of the absolute magnitudes. 

The same holds true for the corresponding distribution function for novalike 
variables (which includes also old novae). In these cases the absolute $V$
magnitudes were taken from \citet{Gilmozzi24} for 42 NLs and from
\citet{Selvelli19} for 17 novae (excluding the long period system GK~Per), 
noting that in the latter case the inclination correction was re-calculated 
using the \citet{Webbink87} prescription. 

The resulting normalized distribution functions are shown in Fig.~\ref{Mv-distr}
for novalike variables in solid black, and as solid red and dashed magenta 
lines for dwarf novae in quiescence and outburst maximum, respectively. They
have distinct maxima, the average values being $M_V = 4.4 \pm 0.9$ (novalike
variables), $M_V = 4.3 \pm 0.8$ (dwarf novae during outburst
maximum) and $M_V = 6.9 \pm 1.1$ (quiescent dwarf novae). It is noteworthy
that the distribution function of dwarf novae during maximum is remarkably
similar to that of novalike variables.

\section{Frequency analysis of some anomalous Z~Cam stars}
\label{Frequency analysis of some anomalous Z Cam stars}

The TESS light curves of some of the anomalous Z~Cam stars discussed in
Sect.~\ref{Relationship to anomalous Z Cam stars} exhibit features in
their power spectra that are either unusual or put into question previous
understandings. Therefore, I briefly present them here without trying to 
investigate them in depth.

\subsection{V513 Cassiopeiae}
\label{V513 Cas}

\begin{figure}
\figurenum{A1}
\label{app-B-1}
\plotone{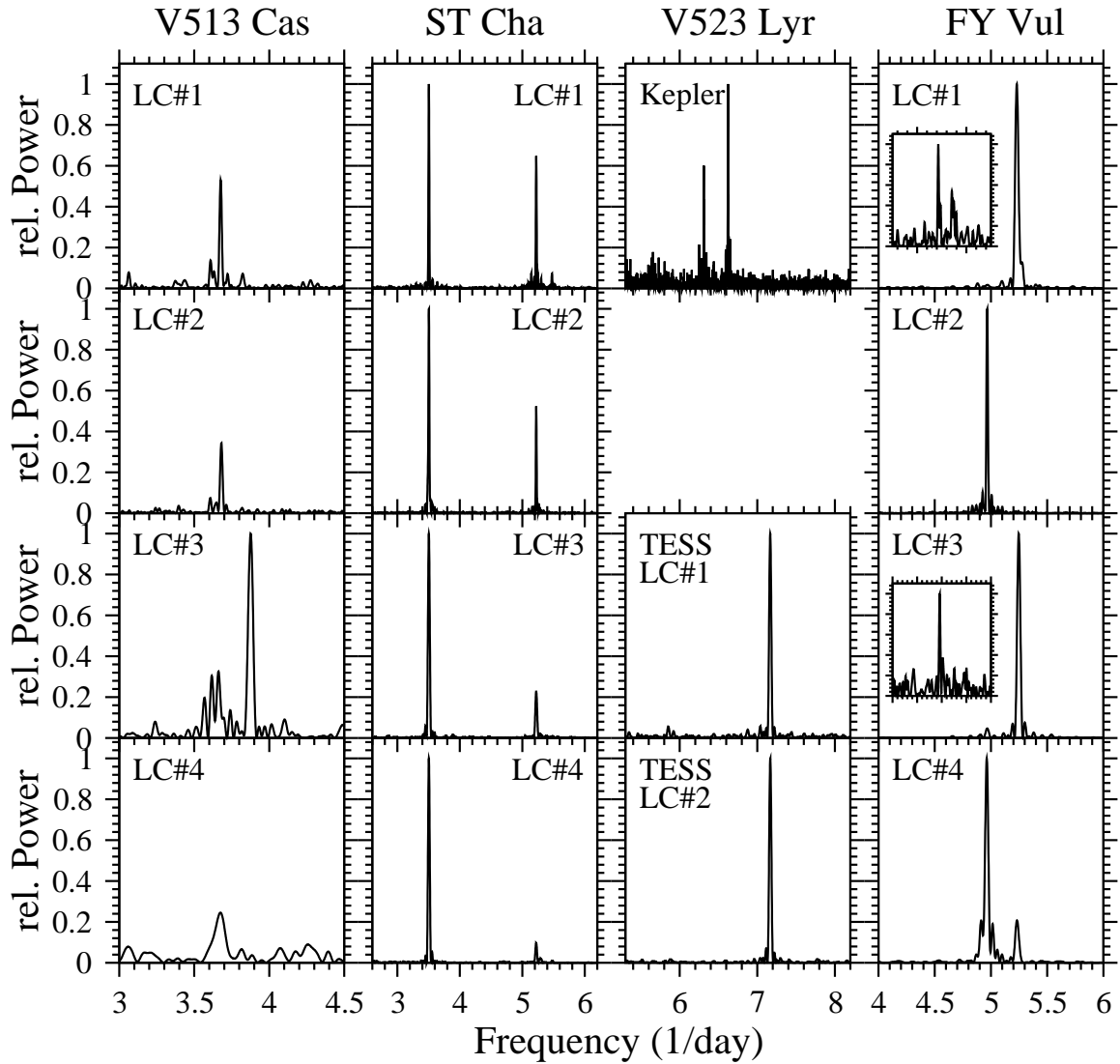}
\caption{Power spectra of TESS and Kepler light curves of four
         systems with special features. For details, see text. The
         frequencies shown in the insets in the right column range
         between 14.5 and 16.5~d$^{-1}$.}
\end{figure}

\citet{Szkody13} determined a spectroscopic orbital period of V513~Cas
of $P_{\rm S} = 312\, {\rm min} = 0.217\, {\rm d}$. The power spectra of 
the four available TESS light curves are displayed in the left column of
Fig.~\ref{app-B-1}. 
In LC\#1 and LC\#2 a signal corresponding to an average period of 
$P_1 = 0.27186(8)\, {\rm d}$ is seen. It is not well defined in
LC\#3 which instead contains a strong signal at $P_2 = 0.2580(2)\, {\rm d}$. 
LC\#4 encompasses only half of a TESS sector. Therefore, the spectral 
features are much broader and are ignored here. Considering that
$1/P_{\rm S} - 1/P_1 = 0.95 \approx 1$ it may well be that the spectroscopic
period of \citet{Szkody13} is a 1/day alias of the true orbital period $P_1$.
$P_2$ is then readily interpreted as being due to a negative superhump.
 
\subsection{V523 Lyrae}
\label{V523 Lyr}

In the Kepler light curve of V523~Lyr \citet{Mason16} identified two 
periodic modulations. The power spectrum of the entire light curve in the
relevant frequency range is shown in the upper panel of the third column
of Fig.~\ref{app-B-1}. 
\citet{Mason16} interpret a persistent variation at 0.15845048~d
as orbital, and a shorter one at 0.151~d, which is only visible during
a short part of the light curve, as being caused by a negative superhump.
None of these signals in seen in the two TESS light curves. Instead,
the power spectra of both light curves have a strong signal at a higher 
frequency (lower panels of the third column of Fig.~\ref{app-B-1}). 
The average corresponds to a period of 0.13953(2)~d. I
have no immediate explanation for the discrepancy between the periods
detected in the Kepler and the TESS data. But
V523~Lyr is a faint star located in a rich field. Keeping in mind the
coarse pixel size of TESS contamination of the light curve by neighboring
stars is therefore well possible.

\subsection{ST Chamaeleontis}
\label{ST Cha}

\citet{Steiner88} first observed a photometric period of 6.85~hr or 0.285~d
in ST~Cha which they interpreted as orbital. This same period is consistently 
present in the power spectra of the four TESS light curves (second column
of Fig.~\ref{app-B-1}), 
permitting to refine it to 0.28532(3)~d. However, all spectra contain 
another significant and stable signal, corresponding to a period of
0.19148(3)~d. The relationship between the two periods is not obvious.
Their large difference precludes to interpret them as a pair of orbital
and superhump variations. ST~Cha is quite bright in a poorly populated 
field so that contamination by neighboring stars is not an issue in this
case.

\subsection{FY Vulpeculae}
\label{FY Vul}

The orbital period of FY~Vul is unkown. The power spectra of four TESS
light curves are shown in the right column of 
Fig.~\ref{app-B-1}. Clearly, two different signals are present. The first
one corresponding to a period of $P_1=0.20139(9)$~d is absent in 
LC\#1, weak in LC\#3 and strong in LC\#2 and LC\#4. The second one with a 
period of $P_2=0.1910(4)$~d is absent in LC\#2, modestly strong in 
LC\#4 and dominating in LC\#1 and LC\#3. It is close at hand to identify 
$P_1 = P_{\rm orb}$ with the orbital period of FY~Vul and $P_2 = P_{\rm nSH}$
with the period of a nSH. Curiously, whenever the superhump is strong a weak
signal is seen corresponding to period of $P_3 = 0.06475(9)$~d 
(insets in the corresponding panels in 
Fig.~\ref{app-B-1}). Another structure to the right of this signal
in LC\#1 can be identified with the second overtone of the superhump
frequency ($P_{\rm nSH}/3$). While $P_3$ is clearly not an overtone
of $P_{\rm orb}$ it is interesting to note that within the error margins the
frequency difference $3/P_{\rm sh} - 1/P_3$ is equal to the frequency
difference $1/P_{\rm nSH} - 1/P_{\rm orb}$. 

\subsection{LAMOST J065237.19+243622.1}
\label{LAMOST J065237.19+243622.1}

\begin{figure}
\figurenum{A2}
\label{app-B-2}
\plotone{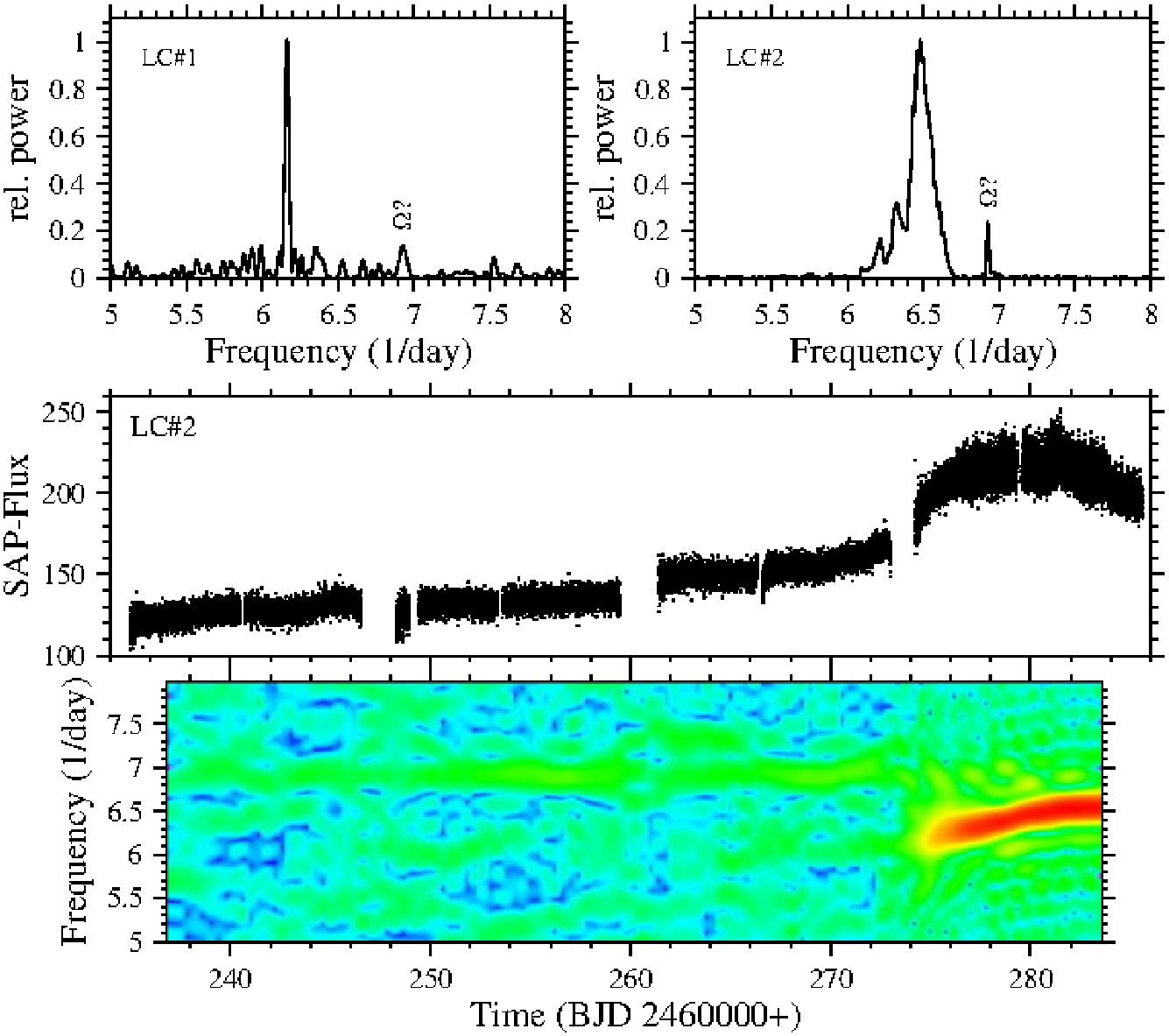}
\caption{{\it Top:} Power spectra of the two TESS light curves of 
         LAMOST~J065237.19+243622.1. The tentatively identified orbital
         signals are marked as $\Omega ?$ (see text). {\it Middle:}
         Light curve LC\#2 of LAMOST~J065237.19+243622.1. {\it Bottom:}
         Time resolved power spectrum of LC\#2 on the same time scale. 
         In order to enhance fainter
         features the power has been scaled to its fourth root.}
\end{figure}

The periods exhibited by LAMOST~J065237.19+243622.1 are confusing. In the
Kepler K2 light curve \citet{Sun24b} identified two periods at 0.15511 and
0.15419~d. The first (LC\#1) of two TESS light curves contains a period of
0.16224(9)~d. \citet{Sun24b} interpret it as orbital and claim the 
presence of a second periodicity at 0.15659~d. The latter period
and those seen in the Kepler data
may indicate a nSH with a somewhat unstable period. The power spectrum is
shown in the upper left panel of Fig.~\ref{app-B-2}. The second TESS light curve
(LC\#2) was not yet available when \citet{Sun24b} submitted their paper. Its
power spectrum is shown in the upper right panel of the figure. 
Surprisingly, it is dominated by a broad feature, encompassing the period
range between 0.150 and 0.164~d, and peaking at 0.1543~d. This range includes
all previously seen periods. The broad feature is flanked by
a narrow signal at 0.14440(3)~d. 

The time resolved power spectrum of LC\#2, using a sliding window of 4~d width,
sheds some light on this issue. It 
is reproduced in the lower panel of Fig.~\ref{app-B-2}, while its middle panel
contains the light curve on the same time scale. The broad feature in the
power spectrum of the whole light curve is
generated by a strong signal with rapidly changing frequency that only appears 
close to the end of the light curve concurrently with the maximum of a
brightening of LAMOST~J065237.19+243622.1. In contrast, the narrow signal,
while weak, is persistent over its whole extend. While
the complex behavior of the variations in the system warrants a deeper
study than can be done here, I permit myself to speculate that the 0.14440~d
period is the true orbital period (noting that in the power spectrum of
LC\#1 an otherwise inconspicuous peak is present at exactly the same
frequency as in LC\#2; see Fig.~\ref{app-B-2}), while all other signals are 
due to a positive superhump with a somewhat variable frequency.

\end{document}